\documentclass[useAMS,usenatbib]{mn2e}
\usepackage{graphicx,subfigure,amsmath,wasysym, amssymb}
\usepackage{setspace}

\title[Hoag's Object Revisited]{Hoag's Object: Evidence for Cold Accretion onto an Elliptical Galaxy}
\author[Ido Finkelman et al.]{Ido Finkelman$^{1}$\thanks{E-mail: ido@wise.tau.ac.il (IF); moisav@gmail.com (AM); noah@wise.tau.ac.il (NB); katkov.igor@gmail.com (IK)}, Alexei Moiseev$^{2}$, Noah Brosch$^{1}$, Ivan Katkov$^{3}$\\
$^{1}$The Wise Observatory and the School of Physics and
Astronomy, the Raymond and  Beverly Sackler Faculty of Exact
Sciences, \\ Tel Aviv University, Tel Aviv 69978, Israel\\
$^{2}$Special Astrophysical Observatory, Russian Academy
of Sciences, Nizhniy Arkhyz, Karachai-Cherkessian Republic,
357147, Russia\\
$^{3}$Sternberg Astronomical Institute, Moscow State University, 13 Universitetski
prospect, 119992, Moscow, Russia}

\date{Accepted 2011 August 8.  Received 2011 August 7; in original form 2011 June 14}

\pagerange{\pageref{firstpage}--\pageref{lastpage}} 

\begin{document}

\maketitle

\label{firstpage}

\begin{abstract}
We present new photometric and spectroscopic observations of the famous Hoag's Object, a peculiar ring galaxy with a central roundish core. 
The nature of Hoag's Object is still under controversial discussion. Previous studies demonstrated that a major accretion event that took place at least 2-3 Gyr ago can account for the observational evidence. However, the role of internal nonlinear mechanisms in forming the outer ring was not yet completely ruled out. 

The observations reported here consist of WFPC2 optical data retrieved from the Hubble
Space Telescope archive as well as long-slit and 3D spectroscopic data obtained at the Russian BTA 6-m telescope. 
These new data, together with HI and optical information from the literature, are used to demonstrate that Hoag's Object is a relatively isolated system surrounded by a luminous quasi-spiral pattern and a massive, low-density HI disc. The main stellar body is an old, mildly triaxial elliptical galaxy with very high angular momentum.

We review previous formation scenarios of Hoag's Object in light of the new data and conclude that the peculiar morphology could not represent a late phase in barred early-type galaxies evolution. In addition, no observational evidence supports late merging events in the evolution of the galaxy, although further tests are required before safely dismissing this idea.
Combining all the information we propose a new scenario where the elliptical core formed in the early Universe with the HI disc forming shortly after the core by prolonged `cold' accretion of primordial gas from the intergalactic medium. 

The low gas density does not allow intense star formation to occur everywhere in the disc, but only along a tightly wound spiral pattern of enhanced density induced by the triaxial gravitational potential. 
According to this view, the physical mechanism that forms rings in Hoag-like galaxies is closely linked with that in some non-barred disc galaxies, although the formation and evolution of both classes of galaxies are clearly distinct. 
Whether or not this unique evolutionary track is related with the galaxy residing in a underdensed environment remains to be solved. 
A detailed HI mapping of Hoag's Object and its environment is required to test our hypothesis and to examine the nature of the HI disc.

\end{abstract}


\section{Introduction}
Hoag's Object (Hoag 1950) has been fascinating professionals and non-professionals astronomers for more than 60 years.
Examining the galaxy image Arthur Hoag described this unique structure as `a perfect halo surrounds a diffuse central image' and suggested that `a proper identification would be a worthwhile short-term project'. 
However, it was only years later that O'Connell, Scargle \& Sargent (1974) attempted a spectroscopic follow up of the object, The authors detected and measured absorption lines of the core but could not trace the emission from the faint ring. Schweizer et al.\ (1987) confirmed by optical spectroscopy and radio observations that the core and the ring have almost identical recession velocities, implying that they are physically associated and at rest relative to each other. These observations also established the extragalactic nature of the galaxy while ruling out a previous hypothesis that the ring was produced by gravitational lensing of a background galaxy (Hoag 1950; O'Connell et al.\ 1974).

For many years the elusive nature of Hoag's Object and its apparent peculiarity among other galaxies were considered to be almost `pathological'. 
O'Connell, Scargle \& Sargent (1974) ruled out the possibility that absorption or scattering by dust particles surrounding the galaxy are responsible for the ring appearance. Alternatively, they suggested that the ring could be the result of an encounter with a companion or a passing galaxy. The formation of the ring through galaxy-galaxy interaction was further studied by Schweizer et al.\ (1987) who found the low relative velocity between the core and the ring to be highly unlikely for a collisional ring galaxy. The authors concluded that Hoag's Object formed by mass transfer or merger during a major accretion event through a mechanism similar to that forming polar ring galaxies. The authors also pointed out that Hoag's Object itself might eventually turned out to be an E0 galaxy with a polar ring. However, several issues remained to be explained in the Schweizer et al.\ (1987) scenario, such as the alignment of bright knots along `a narrower ring which appears off-centre from the main ring'.

A different scenario for the formation of the ring was put forward by Brosch (1985) in the context of weak non-axisymmetric internal perturbations. Brosch suggested that Hoag's Object represents a transient stage in the evolution of a barred galaxy which went through a dramatic bar instability. In this view, the bar in a galactic disc drives part of the gas into an outer ring (Kormendy \& Kennicutt 2004) while triggering intense star formation in the gas-rich disc. 
The dissolved bar is expected to leave an elongated core with discy features. However, Hoag's Object appears to be a pure round spheroid with no trace of a disc and is therefore difficult to understand in the context of orbit resonance theory.

Following its discovery, Hoag's Object became the prototype of a class of galaxies with bright external rings which appear detached from an inner round core, and several studies aimed to determine their relative frequency in the Universe (e.g., Schweizer et al.\ 1987; Wakamatsu 1990). 
These studies agreed that most Hoag-like objects have elongated cores or faint bars, and that they differ in some distinct way from Hoag's Object. 
Genuine Hoag-like objects are therefore unique and rare galaxies even at present (see also Finkelman \& Brosch 2011).

High resolution images of Hoag's Object, obtained with the Hubble Space Telescope (HST)/WFPC2 as part of
the heritage program, now provide an opportunity to view the spectacular structure of the galaxy in unprecedented detail. 
The photometric observations, together with a detailed kinematic study performed with the Russian 6-m telescope, provide new clues to the formation and evolution of this peculiar galaxy.
We shall assume throughout the paper standard cosmology with $H_0$=73 km s$^{-1}$ Mpc$^{-1}$, $\Omega_m=0.27$ and $\Omega_\Lambda=0.73$.
Following Springob et al.\ (2005), we adopt the distance to the galaxy of 175.5 Mpc, which corresponds to a scale of 851 pc
arcsec$^{-1}$.

The paper is organized as follows: Section \ref{S:Obs_and_Red} 
gives a description of the observations and data reduction; we study the object's observed properties in Section \ref{S:analysis} and discuss its structure in Section \ref{S:structure}. In Section \ref{S:hypothesis} we review the old formation hypotheses and suggest our own interpretation of the observational evidence. Our conclusions are summarized in Section \ref{S:conclusions}.

\section{Observations and Data Reduction}
\label{S:Obs_and_Red}
\subsection{HST observations}
Hoag's Object was observed with the HST/WFPC2 for 6 orbits on June 9 2001 under the Hubble Heritage Project. Long exposures were taken in the following filters: $F450W$, $F606W$, and $F814W$ with total exposures of $5\times1300$, $4\times500$ and $5\times1000$ s, respectively. The data used in this study were taken from the WF3 camera, with an image scale of 0.1 arcsec pixel$^{-1}$.
The data were processed by the Space Telescope Science Institute (STScI) pipeline and individual exposures in each filter were combined using the CRREJ task in {\small IRAF}\footnote{{\small IRAF} is distributed by the National Optical Astronomy Observatories (NOAO), which is operated by the Association of Universities, Inc.\ (AURA) under co-operative agreement with the National Science Foundation} for cosmic-ray recognition and rejection. 

\begin{figure*}
\begin{center}
\begin{tabular}{cc}
\includegraphics[width=8cm]{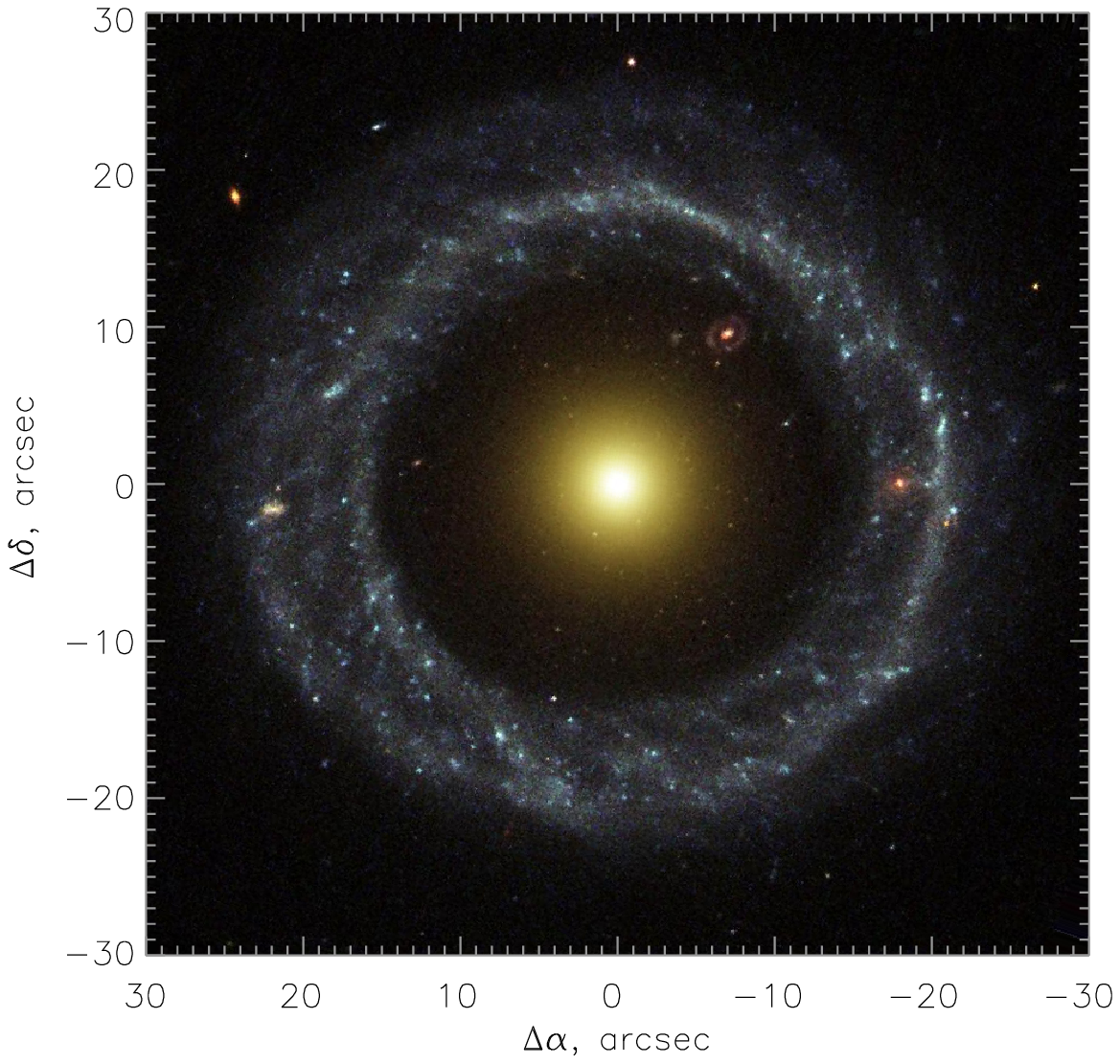} & \includegraphics[width=8cm]{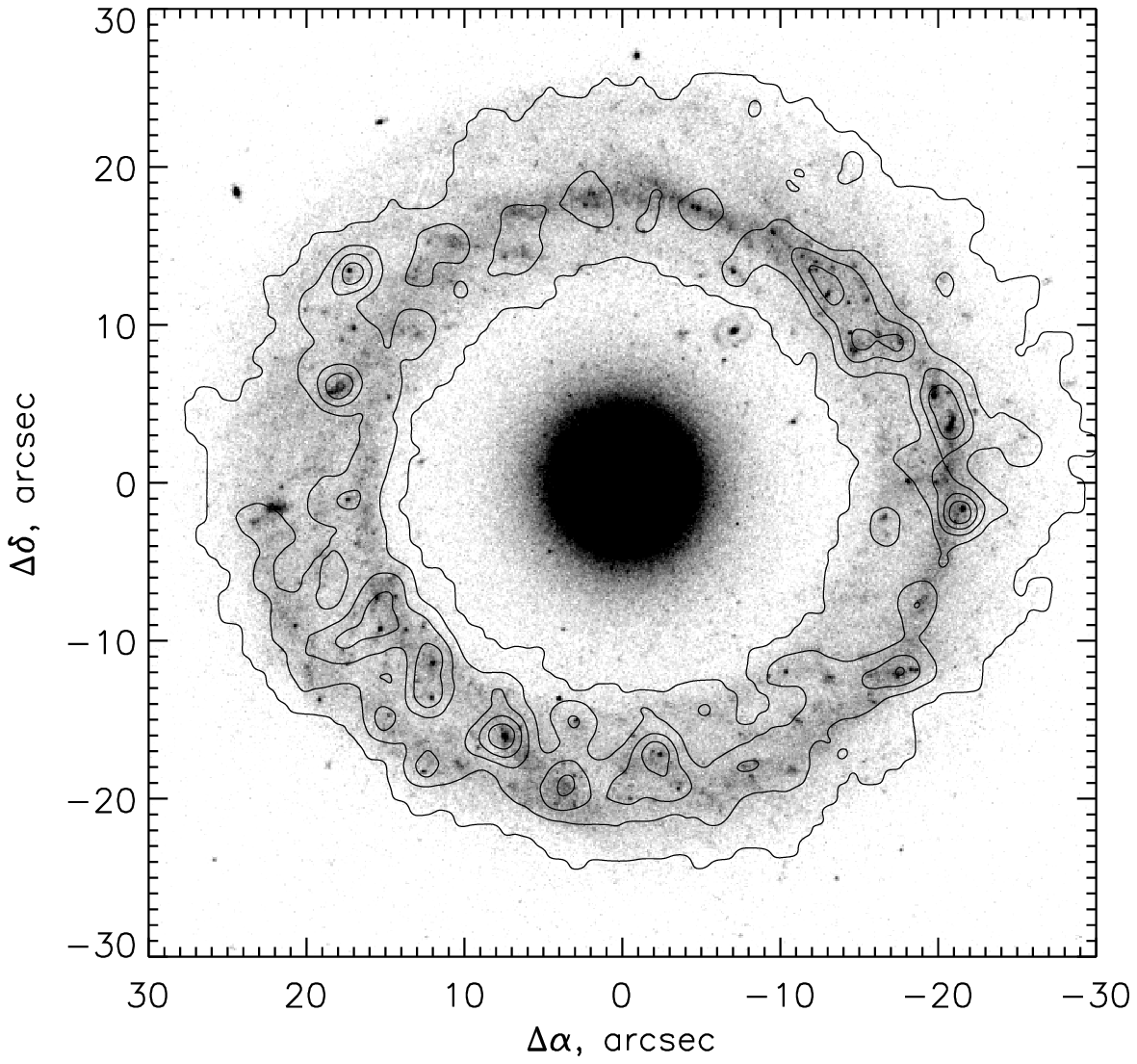}
\end{tabular}
\caption{Left: Hubble Heritage WFPC2 colour image of Hoag's Object (Image credit: NASA, ESA, and the Hubble Heritage Team).
Right: The FPI H$\alpha$ data drawn as contours over the greyscale HST $F606W$ image. \label{fig:HST_color}}
\end{center}
\end{figure*}

To convert the measured instrumental magnitudes to the Vega photometric system we adopted the photometric zero points of 22.016 and 21.659 mag for the $F450W$ and $F814W$ bands, respectively, as given in the WFPC2 photometric cookbook. The $F450W$ and $F814W$ filters are equivalent to the Johnson-Cousins $B$ and $I$ magnitudes, but to allow optimal comparison with previous ground-based observations we transformed our magnitudes to the `standard' $B$- and $I$-bands using the formulae given by Matthews et al. (1999). 
Below we shall refer to the transformed $F450W$ as `$B$' and $F814W$ as `$I$'.
The HST colour image of Hoag's Object, displayed in the left panel of Fig.\ \ref{fig:HST_color}, was obtained from the STScI Web site and is part of the Hubble Heritage Team collection.

\subsection{BTA Observations}
This subsection details the observations made with the 6-m large altazimuthal telescope (BTA) of the Special Astrophysical Observatory of the Russian Academy of Sciences (SAO RAS). The data were obtained with the multi-mode focal reducer SCORPIO (Afanasiev \& Moiseev 2005) and the integral-field Multi-Pupil Fiber Spectrograph (MPFS; Afanasiev, Dodonov \& Moiseev 2001) at the prime focus of the BTA. The MPFS was used to construct a detailed picture of the stellar kinematics in the central region of the galaxy. 
SCORPIO allows various spectroscopic and photometric observations to be performed within a 6 arcmin field of view.
It was operated in a long-slit spectroscopic (LSS) mode to study the large-scale kinematics and stellar population properties, and in a mode of scanning Fabry-P\'{e}rot interferometer (FPI) to study  the ionized gas kinematics in the outer ring. 
In addition, a deep SCORPIO exposure was taken in the Cousins $R$-band to search for faint morphological peculiarities and low-luminosity tidal features around the galaxy.
The detectors used in 2009 and 2010 were a CCD EEV42-40 ($2048\times2048$ pixels) and a CCD EEV42-90 ($4612\times2048$ pixels), correspondingly.
The observation log is presented in Table \ref{tobs}.
\begin{figure*}
\begin{center}
\begin{tabular}{cccc}
{} & \includegraphics[width=8cm]{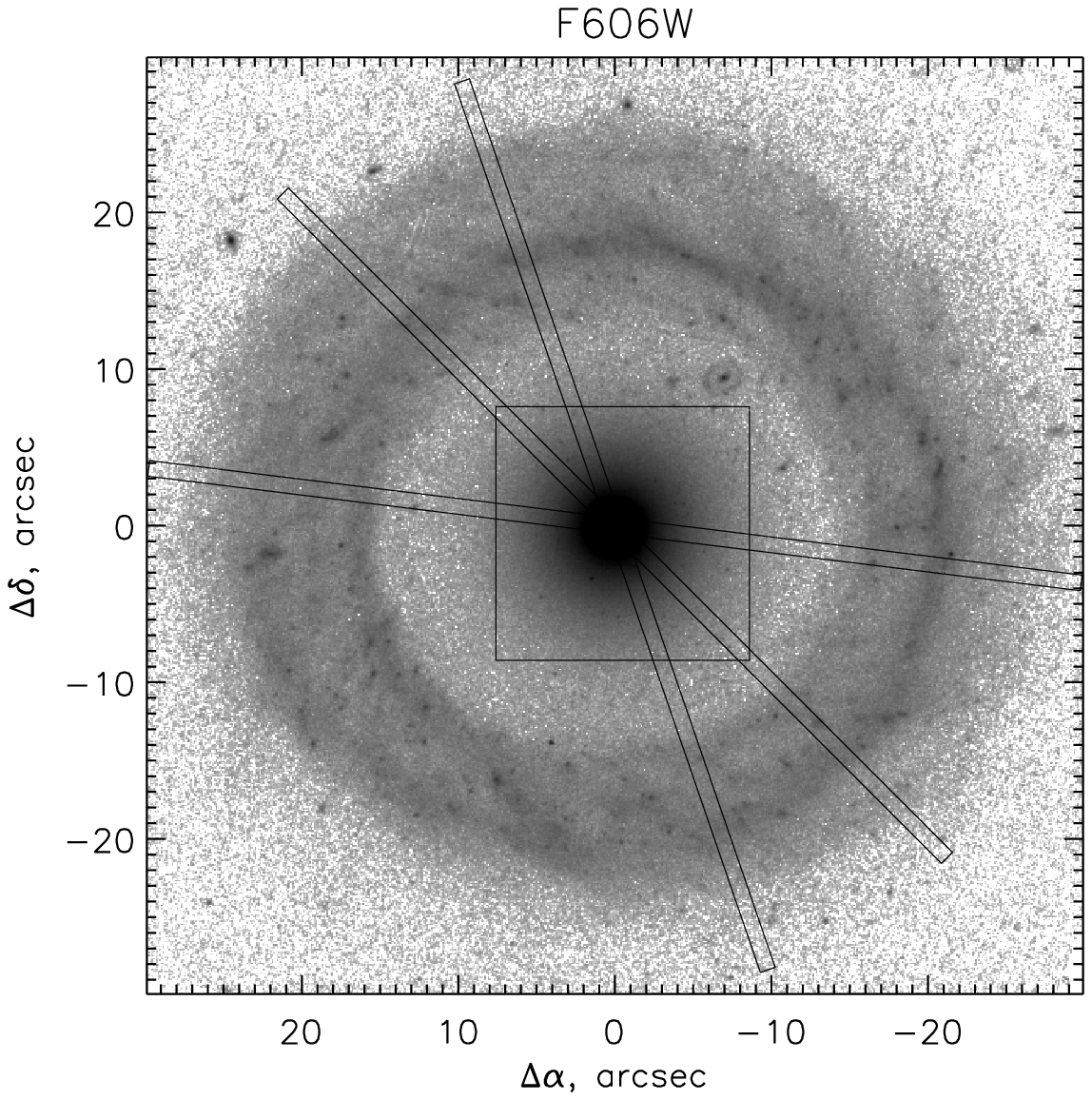} & {} & \includegraphics[width=8cm]{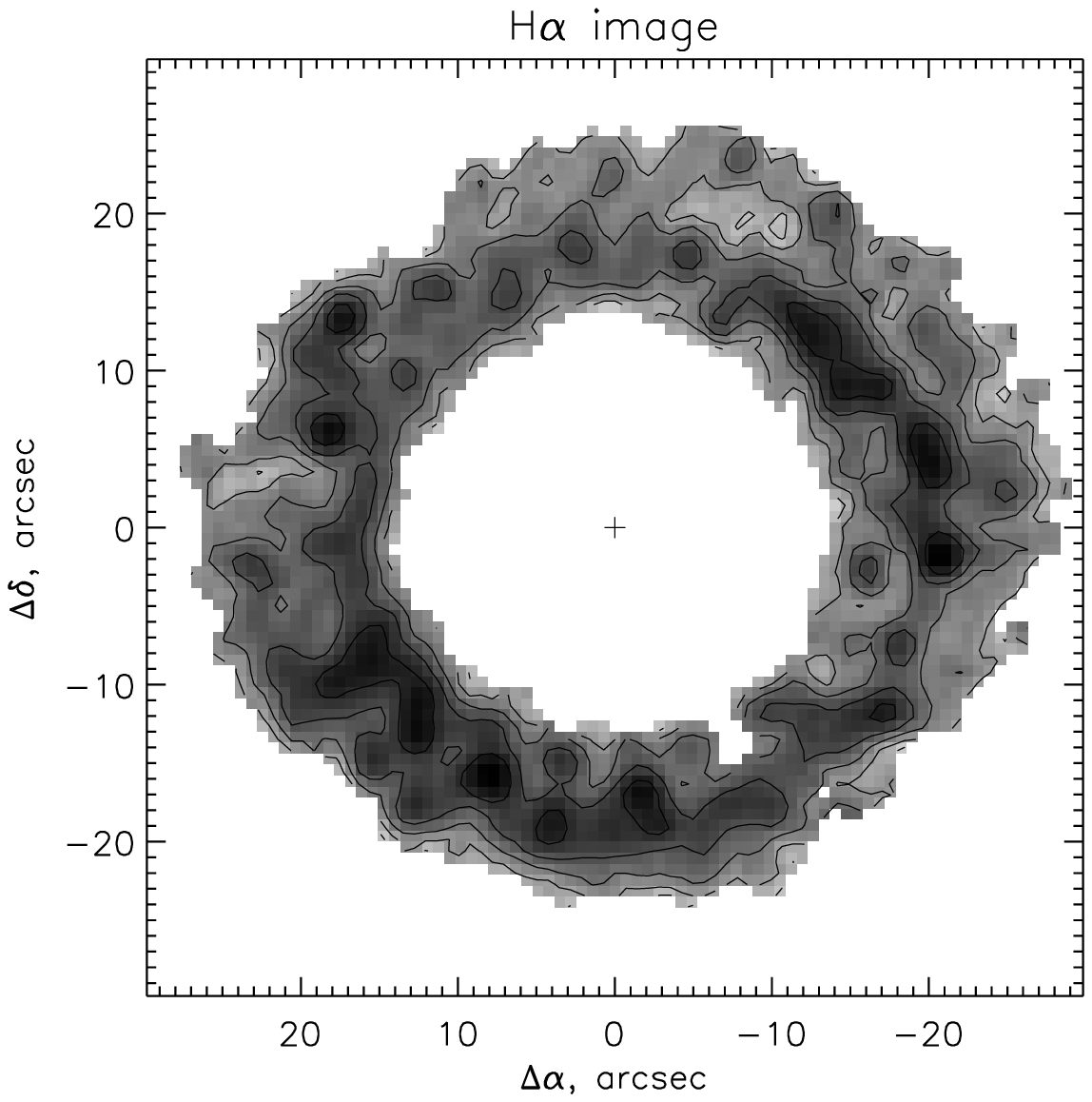}\\
{} & \includegraphics[width=8cm]{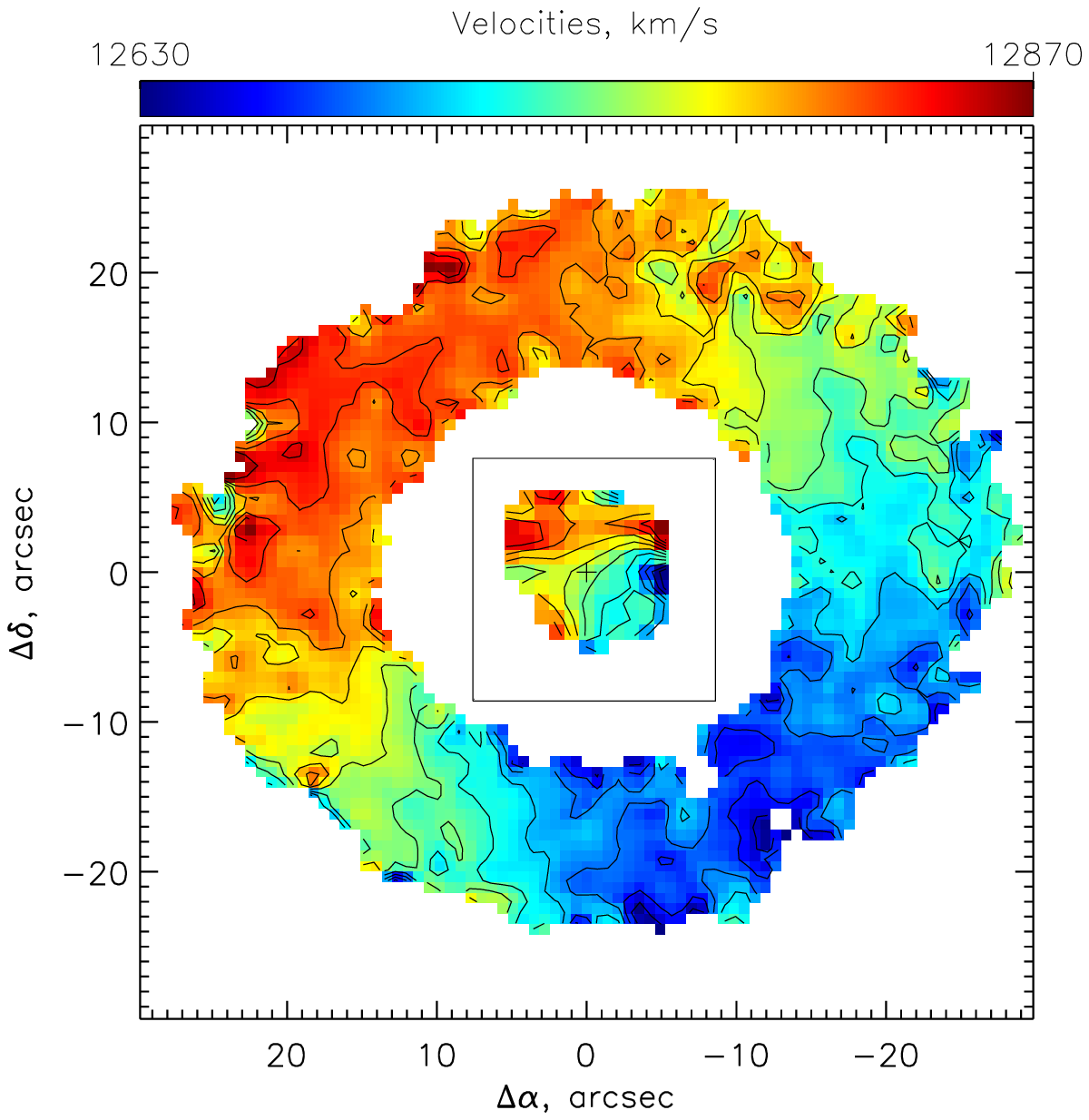} & {} & \includegraphics[width=8cm]{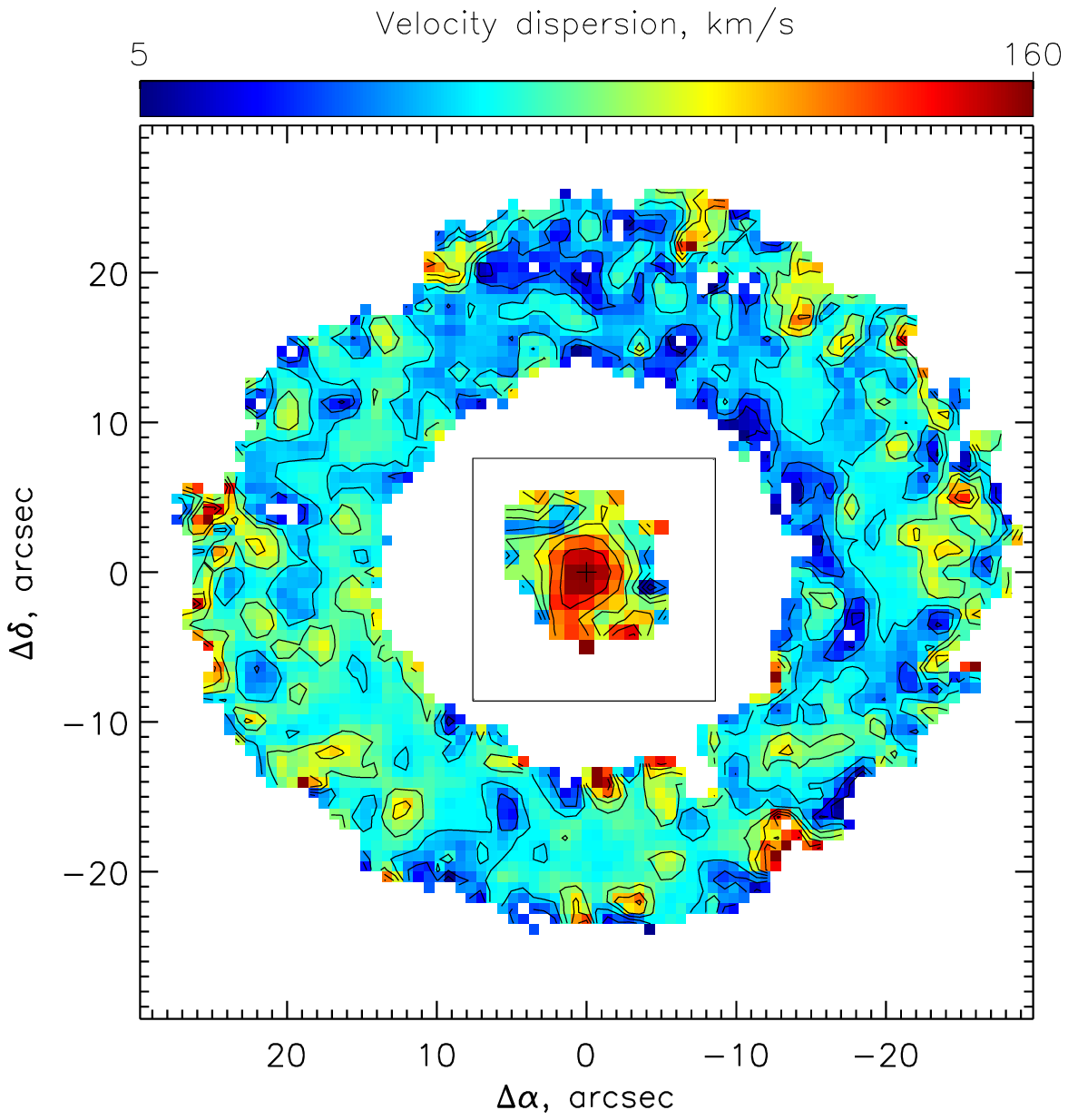}
\end{tabular}
\end{center}
  \caption{Kinematic information for Hoag's Object. Upper panels: WFPC2 $F606W$ image (upper left) and H$\alpha$ intensity map (upper right). Slit orientations and actual slit widths for the three position angles used for spectroscopy are also indicated, as well as the MPFS field of view. Lower panels: The line-of-sight velocity field (lower left) and velocity dispersion field (lower right) of Hoag's Object from FPI and MPFS observations. \label{kinematic_figures}}
\end{figure*}

\subsubsection{MPFS integral field spectroscopy}
The integral-field  spectrograph  MPFS takes simultaneous spectra of 256 spatial elements arranged in a square by a $16\times16$ arcsec$^{2}$ lenslet array, with a scale of  1 arcsec per spaxel. An optical fiber is located behind each lens with its other end packed into the spectrograph slit. The sky background spectrum was simultaneously taken with 17 additional fibers located 4 arcmin away from the object. The wavelength interval included absorption features of the old stellar population around the Mg I triplet.
The preliminary data reduction yields a `data cube', where each fiber/pixel in the $16\times16$ arcsec$^{2}$ field of view is a spectrum (Moiseev et al.\ 2004). To construct the stellar velocity field, we use the cross-correlation technique adapted for MPFS data (Moiseev 2001). Spectra of K-type stars obtained during the same night as Hoag's Object were used as templates for the cross-correlation.

The MPFS throughput is not very efficient compared to SCORPIO, thus the object appears relatively weak. 
The MPFS diagonal field of view is about 22 arcsec which means that the ring is farther away and cannot be sampled.
Therefore, the stellar line-of-sight velocity and velocity dispersion maps could be derived only for the central core of Hoag's Object, which allows determining its kinematic axis.

\subsubsection{Long-slit spectroscopy}
We obtained SCORPIO/LSS data for three positions of the slit using different grisms to provide a wide spectral cover and different resolutions. The locations of the entrance slits are plotted in Fig.\ \ref{kinematic_figures} and the instrumental setup is described in Table \ref{tobs}. 
The slit width was 1 arscec and the spatial sampling was 0.35 arcsec pixel$^{-1}$. 

The position angles (PAs, measured counterclockwise from the north) of the slit were chosen at first to lie along the apparent major axes of the inner spheroid and of the ring 
as found by Schweizer et al.\ (1987), 83$^\circ$ and 19$^\circ$, respectively.
The spectral range covered the H$\alpha$ emission line of the ionized gas and several stellar population absorption features (e.g., Ca I, Fe). 

\begin{figure*}
\begin{center}
\includegraphics[width=18cm]{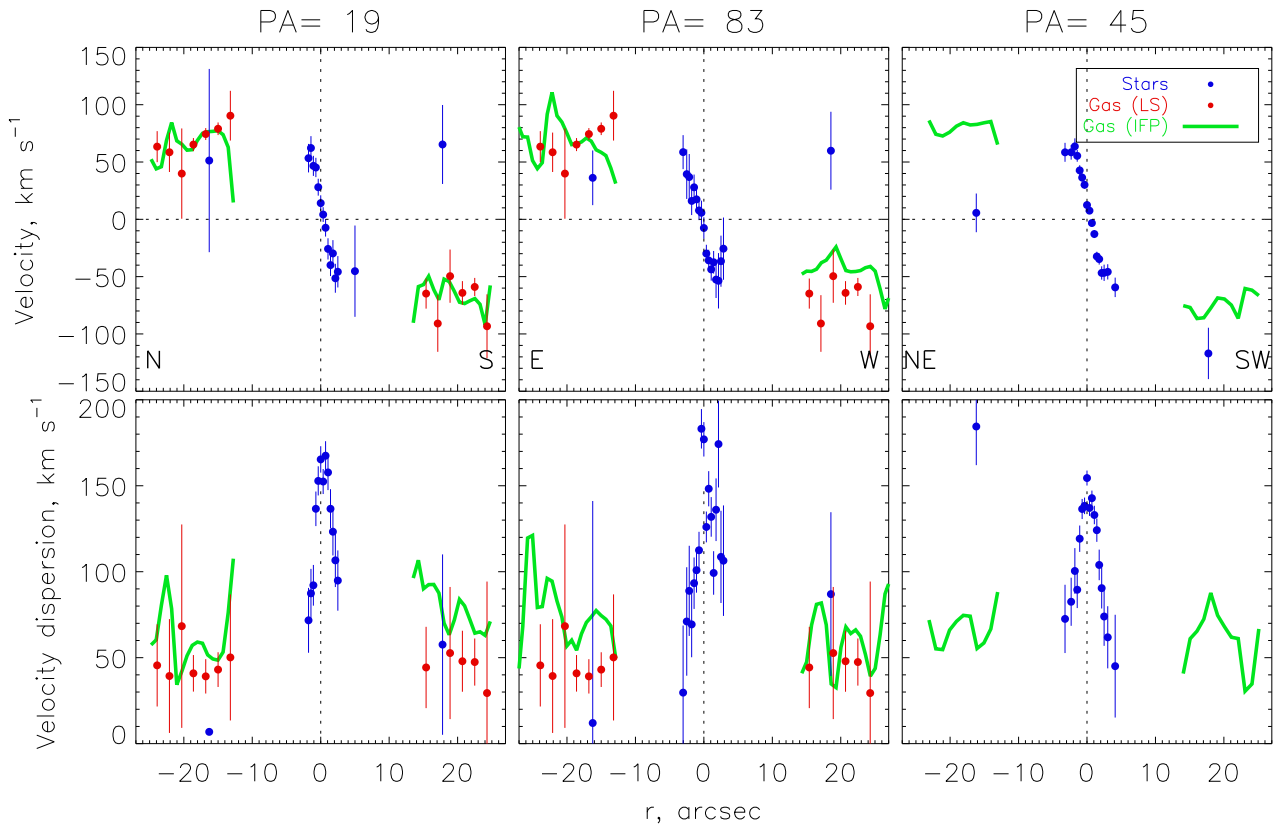}
\end{center}
\caption{The radial distributions of the line-of-sight velocities (systematic velocity was subtracted) and velocity dispersion along different slit positions.
The different symbols show the long-slit spectra measurements of stars and ionized gas; the deepest spectrum was taken along PA=45$^\circ$ in August 2009. 
The solid line represents the pseudo-slit cross-sections taken from the FPI velocity field and velocity dispersion maps.  \label{fig:Hoag_radial}}
\end{figure*}

However, an analysis of the spectroscopic data showed that the kinematics major axes of both the core and the ring coincide and lie at about $45^\circ$, in contrast to the Schweizer et al.\ (1987) results. Observations during May 2009 aimed to obtain deep exposures along the true kinematic axis were conducted under strong cirrus clouds and bad weather conditions. We estimate the total effective exposure time to be $\sim20$ min. Finally, the spectra along PA$\sim$45$^\circ$ were obtained also in August 2009 in a spectral range covering the ionized gas emission lines (e.g., H$\beta$, [OIII]) and the high-contrast stellar absorptions (e.g., Mg I, Fe).
Fig.\ \ref{fig:Hoag_radial} shows the radial distributions of the line-of-sight velocities and velocity dispersion along the different slit positions.

The slit spectra were reduced and calibrated using the IDL-based software, developed at the SAO RAS. 
The parameters of the emission lines (FWHM, flux, velocity) were calculated using a single-Gaussian fitting.

\begin{table*}
\caption{Observation log.} \label{tobs}
\begin{tabular}{@{}clrrlrr}
\hline
Date                 &   Instrument    & PA         & Exposure       & Spectral     & Spectral         & seeing    \\
                     &                 & ($^\circ$) & time (s)       & range (\AA)  &resolution (\AA)  & (arcsec) \\
\hline
27.04.2009           &   MPFS          &   0       &  6000    & $4750-6230$    &     3.5      &    2 \\ 
05.04.2009           &  SCORPIO/LS     &   19      &  7200    & $6100-7050$    &     2.5      &    1.5       \\
05.04.2009           &   SCORPIO/LS    &   83      &  3300    & $6100-7050$    &     2.5      &    1.5      \\
21.05.2009           &   SCORPIO/LS    &   45      &  4800    & $4830-5600$    &     2.2      &    2       \\
14.08.2009           &   SCORPIO/LS    &   45      &  6000    & $4050-5830$    &     5        &    1.3     \\
20.03.2010           &  SCORPIO/FPI    &           &  9600    & H$\alpha$      &     3.4      &    1.7      \\
23.03.2010           &  SCORPIO/IM     &           &  1920    & $R$            &              &    1.4       \\
\hline
\end{tabular}
\end{table*}

\subsubsection{Scanning Fabry-P\'{e}rot Interferometer}
The last set of spectroscopic observations was performed with the scanning FPI mounted on the SCORPIO focal reducer. To cover the spectral range in the region of the redshifted H$\alpha$ line we used a narrow-band (FWHM=20\AA) filter. The free spectral interval between the neighbouring interference orders was 31{\AA} (1420 km s$^{-1}$). The FPI resolution, defined as the FWHM of the instrumental profile, was about 3.4{\AA}  (150 km s$^{-1}$) for a 0.97-{\AA} sampling. The detector was operated in on-chip binned $4\times4$ pixel mode to reduce readout time and match the resultant pixel to the
seeing size. The final image scale and field of view were 0.7 arcsec per pixel and $6.1\times6.1$ arcmin$^2$, respectively. We collected 32 successive interferograms of the object with
different spacings of the FPI plates. A detailed description of the technique of observations and data reduction within IDL-based software can be found in Moiseev (2002) and Moiseev \& Egorov (2008). Following the primary data reduction, the frames were combined into a data cube where each pixel in the  $512\times512$ field contains a 32-channel
spectrum centred on the H$\alpha$ emission line. The final angular resolution after optimal smoothing is about 2.1 arcsec.

Images of the galaxy in the H$\alpha$ emission line and in the $F606W$ filter are shown in the right panel of Fig.\ \ref{fig:HST_color} and in the upper panels of Fig.\ \ref{kinematic_figures}.
The line-of-sight velocity field and velocity dispersion field of the ionized gas were mapped using Gaussian fitting of the emission-line profiles and are presented along with the MPFS velocity maps in the lower panels of Fig.\ \ref{kinematic_figures}. These results are consistent with the radial velocity measurements along the different slit positions, as demonstrated in Fig.\ \ref{fig:Hoag_radial}.

\subsubsection{Imaging}
\begin{figure}
\includegraphics[width=8cm]{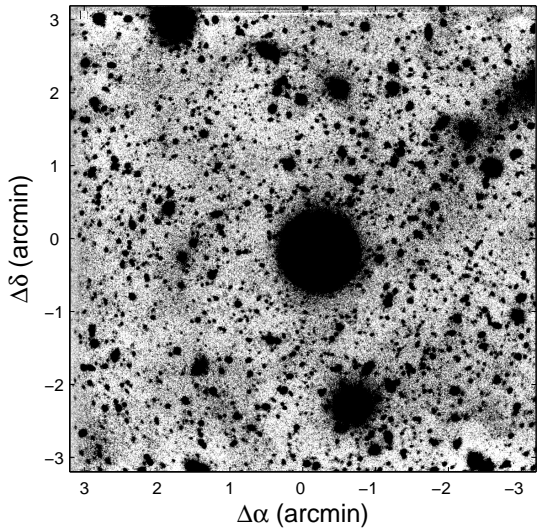} 
  \caption{$R$-band negative image of Hoag's Object obtained with the BTA 6-m telescope. The limiting surface brightness reaches $\mu_R\approx27.3$ mag arcsec$^{−2}$. \label{fig:deep_BTA}}
\end{figure}

We obtained deep exposures of the galaxy with SCORPIO in the Cousins $R$ band with a sampling of 0.351 arcsec pixel$^{-1}$ in the 6 arcmin field of view. The exposures were then aligned and combined into one single 32 min image which is presented in Fig.\ \ref{fig:deep_BTA}. Non-photometric weather conditions prevented the use of standard stars to calibrate
fluxes. To perform a coarse calibration we also observed Hoag's Object with the Wise Observatory (WO) 1-m telescope.
The WO $R$-band observations allowed estimating the zero-point magnitude of the SCORPIO image with an accuracy of $\sim$0.1 mag. The depth of the surface brightness measurements reaches $\mu_R\approx27.3$ mag arcsec$^{-2}$, significantly deeper than other known images of Hoag's Object. 

To calibrate the FPI H$\alpha$ emission flux we obtained with SCORPIO a 900-s exposure with a narrow-band (FWHM=20\AA) filter centred on the emission line and a 240-s exposure with an `off-line' narrow-band (FWHM=50\AA) filter. The calibration was performed by observing the spectrophotometric standard star BD+33d2642 at a similar airmass.

\section{Results}
\label{S:analysis}
\subsection{Environment}
Hoag's Object does not lie within the boundary of a recognized cluster of galaxies (O'Connell, Scargle \& Sargent 1974).
We checked the neighbourhood of Hoag's Object by searching the NASA/IPAC Extragalactic Database (NED) for nearby galaxies with known redshift. 
We found that the nearest galaxy, CGCG 135-050 (a member of a compact group), lies at a linear distance of about 3 Mpc from Hoag's Object, characterizing Hoag's Object as a relatively isolated galaxy (see Spector \& Brosch 2010).

\subsection{Gas content}
The gas content of Hoag's Object was investigated in several studies. 
Brosch (1985) observed the galaxy with the Westerbrock Synthesis Radio Telescope (WSRT) but could not detect any significant line emission at the position of Hoag's Object or elsewhere in the HI maps. Brosch concluded that the total neutral hydrogen content of this object is less than $2.3 \times 10^9 \, \mbox{M}_\odot$. However, observing with the Arecibo 305-m telescope, Schweizer et al.\ (1987) found explicit evidence for the presence of a massive HI content. The authors reported that the 21-cm emission line profile exhibits a two-horned shape typical of rotating discs with a width of 239 km s$^{-1}$. 
The estimated HI mass of $8.2\times10^9 \, \mbox{M}_\odot$ is about 4 times higher than the upper limit estimated by Brosch (1985). 
Since the synthesized beam size of the WSRT observations was about 12 arcsec, while that of the Arecibo telescope was about 3.3 arcmin, the apparent discrepancy between the results could be explained if the neutral hydrogen is diffusely distributed in a disc which extends beyond the stellar core.

Another hint for the extended HI structure of Hoag's Object is provided by the Hoag-type galaxy UGC 4599 which contains a similar amount of HI gas spread over a $\sim$100 kpc disc.  
This implies that genuine Hoag-type galaxies belongs to a class of galaxies with massive, extended HI discs. These structures can extend up to $\sim$200 kpc and are typically characterized by a low surface density $\lesssim$2$\, \mbox{M}_\odot$ pc$^{-2}$ (Sadler, Oosterloo \& Morganti 2002).    

Hoag's Object was included in a search for CO emission in ring galaxies by Horellou et al.\ (1995).
The authors could not detect CO in Hoag's Object and put an upper limit of $7\times10^8 \, \mbox{M}_\odot$ on the H$_2$ mass. 
They suggested that most of the molecular cloud would be in the ring and therefore difficult to detect.
  
\subsection{Surface photometry using the HST}
The HST imaging data were used to provide the morphological parameters, integrated magnitudes and colours of the central core and of the ring of Hoag's Object. 
In order to accurately measure the light of the galaxy, the following technique for the determination of the sky background
was used. For each filter we measured fluxes within a series of concentric ring-shaped apertures of 2 pixel width (the point spread function FWHM) around the object centre. The mean flux of each annulus was then plotted against the radius, which allowed estimating the sky background at the location where the plot decreased to a nearly constant value. 

\begin{figure}
\begin{center}
\begin{tabular}{c}
\includegraphics[trim = 0mm 10mm 0mm 10mm, width=8cm]{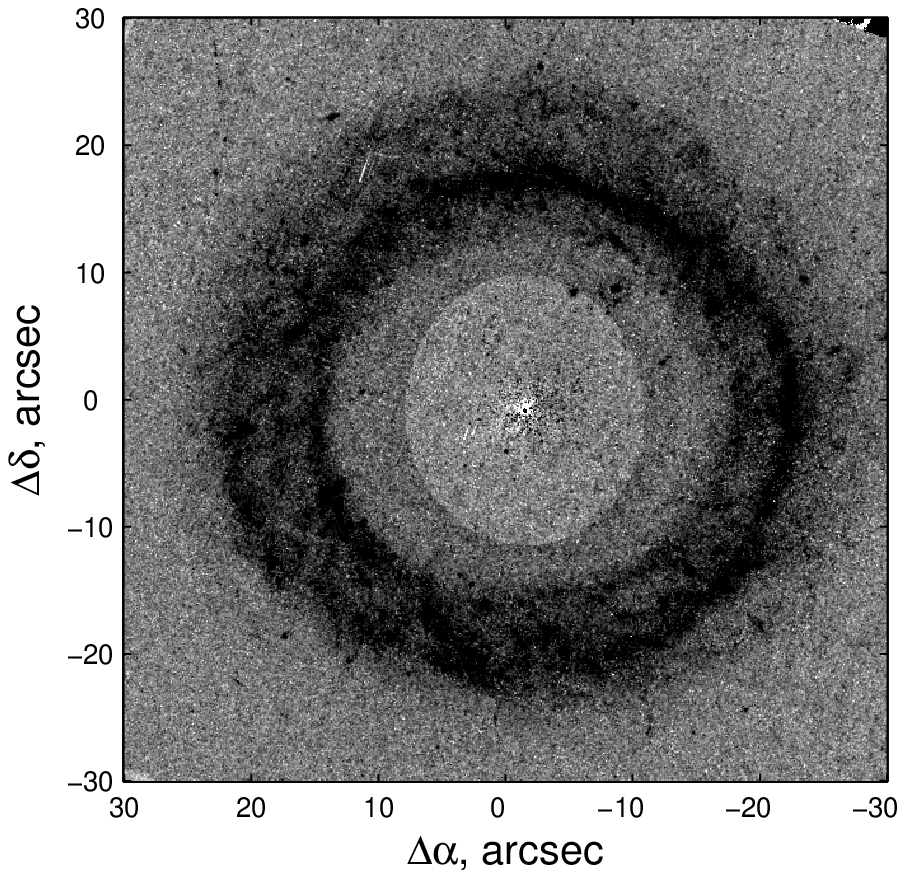} \vspace{8mm} \\
\includegraphics[trim = 0mm 0mm 0mm 0mm, width=8cm]{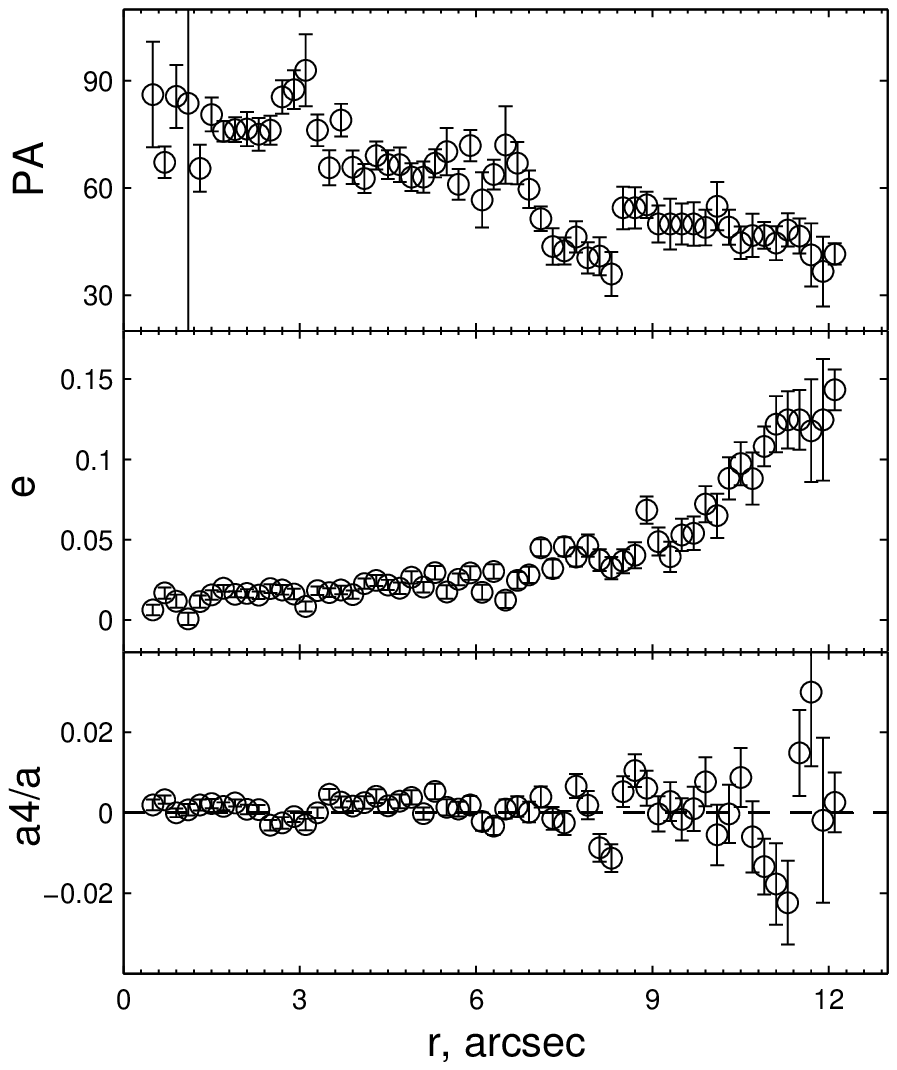} 
\end{tabular}
\end{center}
  \caption{The core of Hoag's Object. Top: Residual image of Hoag's Object. This image was produced by fitting ellipses to the core and subtracting the model from the combined $F450W$ image. The core shows no underlying structure. Bottom: The geometrical properties of the core derived by isophotal ellipse fitting. 
  \label{fig:Hoag_res_bulge}}
\end{figure}

To separate the central core and ring, we used the $F450W$ and $F814W$ images to draw isophotes of the galaxy. This was done by measuring the mean surface brightness of isophotes using the ELLIPSE task in {\small IRAF} by fitting the PA and ellipticity, but keeping the centre fixed. Prominent point sources were masked out to ensure their exclusion from the isophote fit.
Due to the low surface brightness of the intermediate region between the core and the ring, and since the ring exhibits a knotty and non-uniform structure, we restricted the ellipse isophote fit procedure to the luminous core only. 
The ellipticity of the core increases smoothly from $\sim$0.01 at the centre to  $\sim$0.15 at $r=12$ arcsec, whereas the PA profile decreases from $\sim$75$^{\circ}$ to $\sim$45$^{\circ}$ and the isophotal shape parameter ($a_4$/$a$, see e.g., Milvang-Jensen \& Jorgensen 1999) profile lies very close to 0.
The fitting isophotal parameters were fed into the BMODEL task to create a model of the core, which was then subtracted from the actual image to produce the residual image shown in Fig.\ \ref{fig:Hoag_res_bulge}. The figure also presents the radial profiles of the ellipticity, PA and $a_4$/$a$ derived from the ellipse fitting procedure. 

\begin{figure}
\begin{center}
\includegraphics[trim = 0mm 0mm 0mm 0mm, width=8cm]{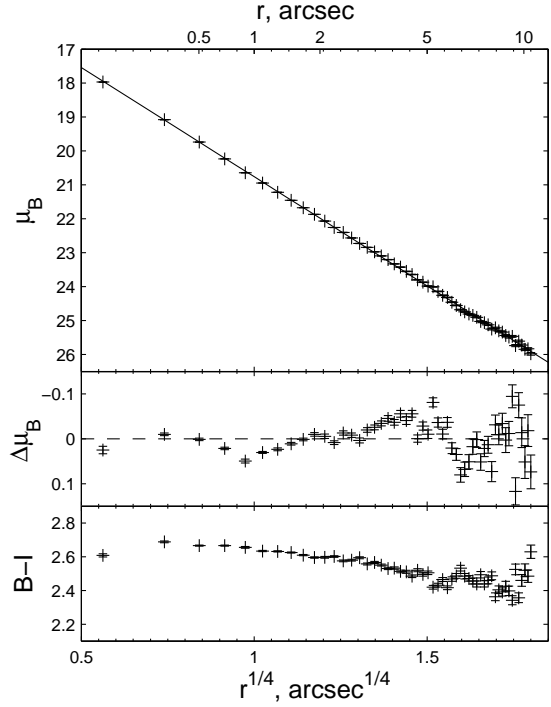}  
\end{center}
  \caption{Surface $B$-magnitude and colour profiles derived by fitting ellipses to the isophotes of the core. The straight solid line on the plots vs.\ $r^\frac{1}{4}$ represents the best-fit to the de Vaucouleurs light profile. \label{fig:Hoag_prop}}
\end{figure}

To examine the structure of the core we fitted the $B$-band surface brightness profile with a two component model consisting of an exponential disc and a S\'{e}rsic bulge. Our least-squares algorithm indicates that the surface brightness profile can be well-fitted by the single `$r^{\frac{1}{n}}$' component over more than $\sim$8 mag range. The fit yields a S\'{e}rsic index, effective radius and effective surface magnitude of $n=3.9\pm0.2$, $r_e=2.8\pm0.1$ arcsec (=$2.5\pm0.1$ kpc) and $\mu_{B}\sim22.6\pm0.5$ mag arcsec$^{-2}$. The fit values are consistent with the de Vaucouleurs light distribution found by Schweizer et al.\ (1987). We plot in Fig.\ \ref{fig:Hoag_prop} the isophotal surface magnitude and colour along the core as measured by the ellipse fitting procedure. As shown in the figure, the de Vaucouleurs `$r^{\frac{1}{4}}$'-law, represented by a solid line, fits well the light profile with slight deviations of less than 0.1 mag. 

\begin{figure}
\begin{center}
\includegraphics[width=8cm]{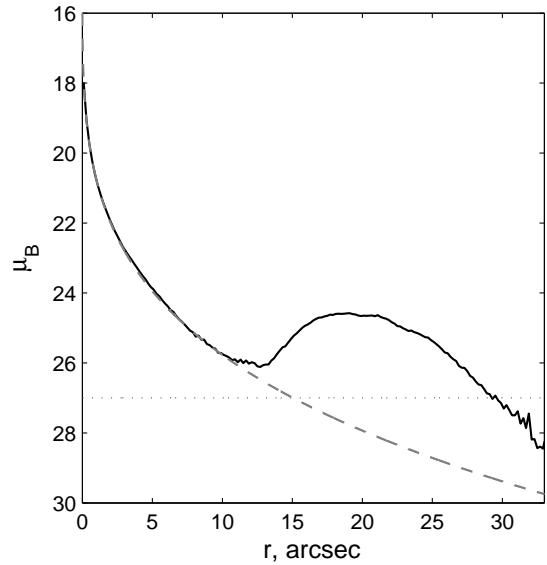} 
\end{center}
  \caption{Azimuthally averaged luminosity profile of Hoag's Object. The dashed line represents an extrapolation of the $r^{\frac{1}{4}}$ light profile to the outer parts of the galaxy. The dashed line represents the 3-$\sigma$ background noise level. \label{fig:Hoag_light_profile}}
\end{figure}

To investigate the light profile outwards of the core we derived an azimuthally averaged luminosity profile (AALP). 
This was done by measuring the intensity within circular apertures centred on the host galaxy photocentre, with radii increasing outwards using ELLIPSE. Since the ellipticity parameter in ELLIPSE cannot be pre-fixed to zero, light profiles are obtained for two orthogonal PAs with minimal ellipticity and their average values are adopted for the AALP.
The AALP, shown in Fig.\ \ref{fig:Hoag_light_profile}, shows a clear distinction between the core and the ring. The core luminosity decreases smoothly from the centre outwards to its faintest level of $\mu_{B}=26.0$ mag arcsec$^{-2}$ at a radius of $r=13.1$ arcsec. The total luminosity of the core is $m_{B}=16.98\pm0.01$ and $m_{I}=14.47\pm0.01$. 
Further away the surface brightness peaks  at a radius of $r=19.1$ arcsec with $\mu_{B}=24.6$ mag arcsec$^{-2}$, although the surface brightness of the ring is rather flat in the range $r\simeq17.0$ arcsec to $r\simeq21.5$ arcsec, where it varies by $\sim$0.1 mag arcsec$^{-2}$. The AALP eventually decreases below $\mu_{B}\sim27.0$ mag arcsec$^{-2}$ at a radius of $r\simeq29.0$ arcsec,  where it drops below the 3-$\sigma$ background noise and the circular aperture approaches the CCD edges.
The ring does not show a sharp outer boundary, and its total luminosity measured out to $r=29$ arcsec is $m_{B}=16.94\pm0.01$ and $m_{I}=15.25\pm0.01$. 

\begin{figure*}
\begin{tabular}{cc}
\includegraphics[trim = 0mm 0mm 0mm 0mm, width=8cm]{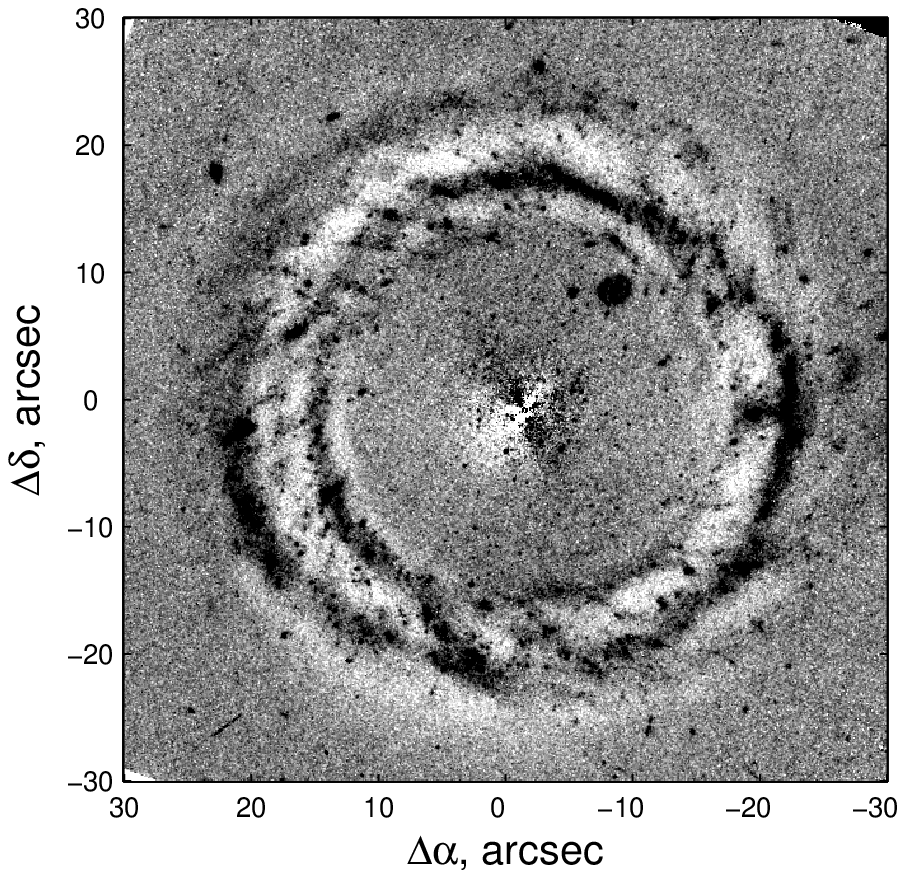} & \includegraphics[trim = 0mm 0mm 0mm 0mm, width=8cm]{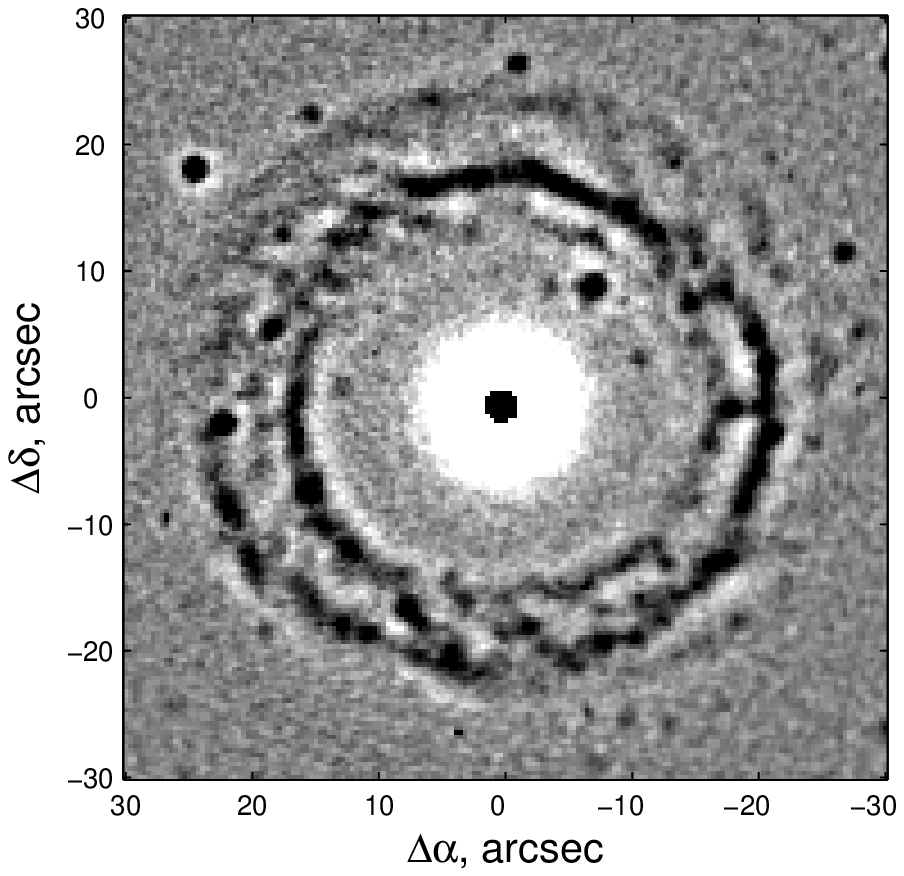}
\end{tabular}
\caption{Left: a residual image of Hoag's Object produced by subtracting the AALP model from the combined HST image. Right: a sharped-masked image obtained by subtracting the median filtered image from the original BTA image. Both images are displayed as negatives.  \label{fig:Hoag_res_all}}
\end{figure*}

To produce a significantly deeper optical image than the individual HST images, the $F450W$-, $F606W$- and $F814W$-band images were co-added. Although the resultant image cannot be adequately flux-calibrated, it is useful to examine the morphology of the galaxy. To enhance fine structures along the ring we created a model for the AALP of the galaxy using the ELLIPSE and BMODEL tasks and subtracted the model image from the original image.
The resultant image, presented in the left panel of Fig.\ \ref{fig:Hoag_res_all}, shows a projected shape of an inner ring and an outer quasi-spiral structure. The tight spiral pattern can also be interpreted as a helix-like structure wound around the central body projected on the plane of the sky. 
This pattern is well-seen also in the sharped-masked image on the right panel of Fig.\ \ref{fig:Hoag_res_all} obtained by subtracting the median filtered image from the original deep BTA image. 

\subsection{Stellar population}
\label{S:pop}
To study the stellar population of Hoag's Object we fitted the obtained LSS spectra with spectra of model stellar populations. To produce the model spectra we used the PEGASE.HR high-resolution spectra which were computed from the ELODIE 3.1 high resolution stellar spectra library for various  ages and metallicities of a single-age stellar population (SSP), assuming a Salpeter (1955) mass function. 
The model spectra were then convolved with the SCORPIO Line Spread Function (LSF) which was derived from twilight sky spectra observed on the same nights as the galaxy. The LSF parameters were estimated over several segments of the wavelength range by fitting the observed spectrum by a broadened high-resolution spectrum of the Sun using the {\tt ULySS} software (http://ulyss.univ-lyon1.fr/). 

The kinematic and stellar population parameters were fitted simultaneously to the spectral observations using {\tt ULySS} via a least-squares minimization technique. 
The emission spectra were then obtained by subtracting the best-fitting model spectra from the observed spectra.
 
The central core spectra do not show any emission lines, and most emission lines in the ring spectra are very weak or undetectable. In fact, only the Balmer emission lines, H$\alpha$ and H$\beta$, could be detected in the `red' ($\sim$6100-7050\AA) and `green' ($\sim$4830-5600\AA) spectral ranges, respectively. Therefore, we were unable to measure the radial distribution of chemical abundance and ionization properties along the ionized gas ring, but only to estimate its kinematic parameters. Instead, the stellar properties and metallicities were derived from the integrated spectrum of the ring.

\begin{figure*}
\includegraphics[]{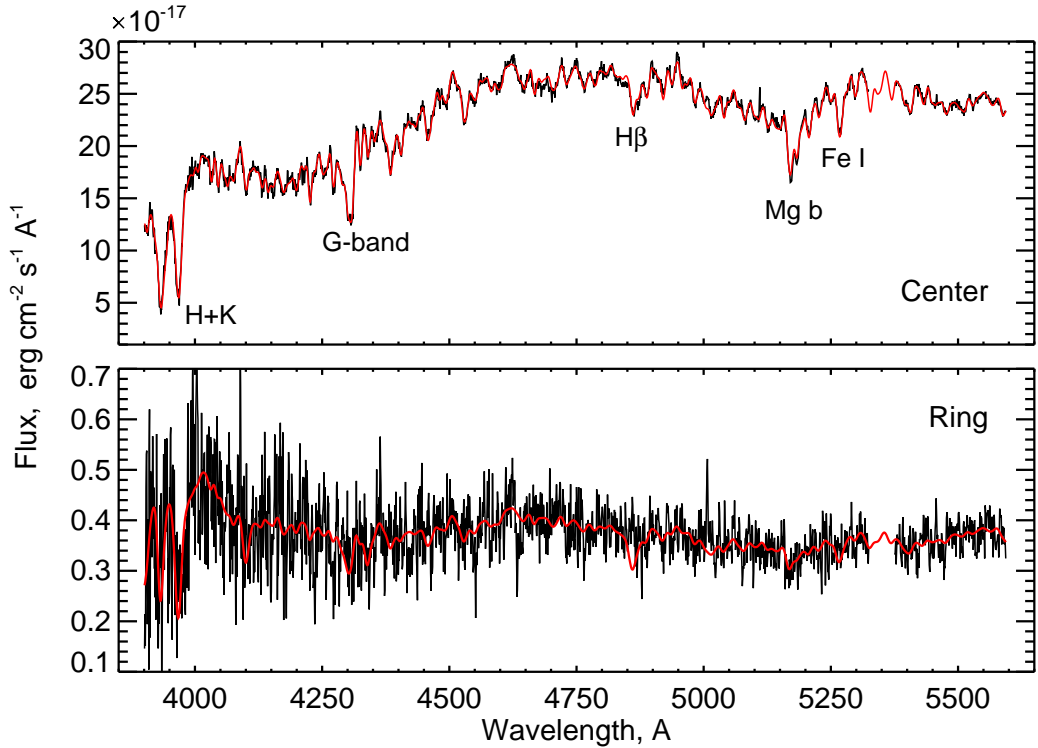} 
\caption{Representative spectra of the core and the ring (black lines) and their best-fit models (red lines).  The main absorption features are marked in the upper panel. The x-axis shows the rest-frame wavelengths.  \label{f:spectra}}
\end{figure*}

\begin{figure*}
\hspace{-1cm} 
\includegraphics[]{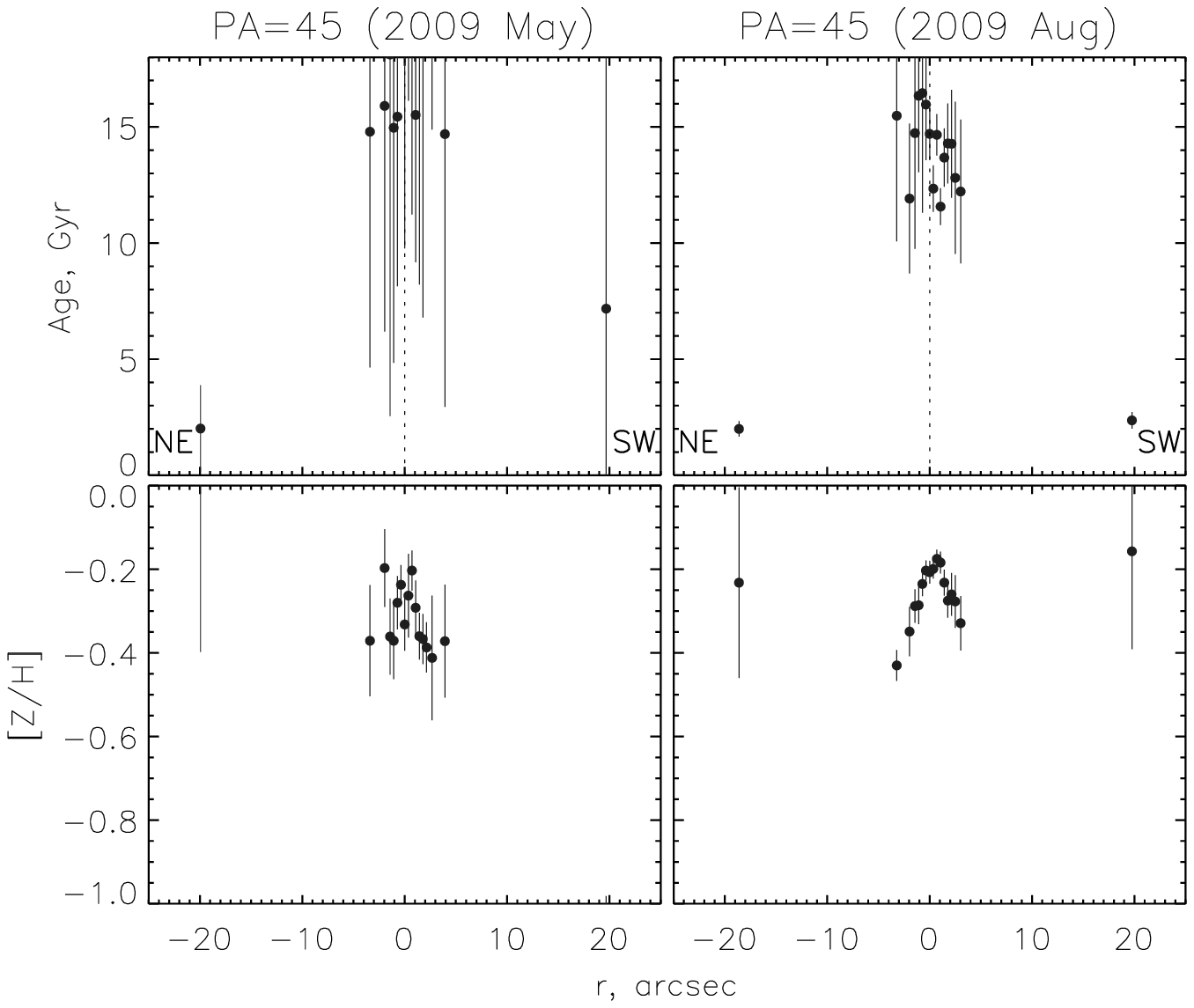}
\caption{The radial distributions of the stellar age and metallicity along PA=$45^\circ$. The May 2009 observational setup had higher spectral resolution but shorter exposure time with respect to the August 2009 setup. \label{fig:age}}
\end{figure*}

Since the `red' spectral range contains relatively low-contrast stellar absorptions, we rely on the `green' domain spectra (taken just along the true kinematic axis) for studying the stellar properties. To fit the data we use most of the absorption lines in the spectral range, including Mg I and Fe (but not H$\beta$). 
Representative spectra of the core and the ring and their best-fit models are presented in Fig.\ \ref{f:spectra}. 
Radial distributions of the stellar age and the metallicity are presented in Fig.\ \ref{fig:age}. The errors of the obtained parameters were estimated by least square minimization.

As shown by these figures, the spectrum of the central core is well-matched by a single $\gtrsim$10 Gyr old stellar population. 
The spectral analysis of the ring indicates the presence of a $\sim$2 Gyr old stellar population.
However, the true light-weighted mean age of stars in the ring is likely considerably smaller than our estimate given that there are clear signatures of ongoing star formation in the ring, and considering that absorption features are not very useful in constraining the age of very young stars.
As can be seen clearly in Fig.\ \ref{fig:HST_color}, where we draw H$\alpha$ contours over the greyscale HST image, the ring is dotted with tens of blue massive star clusters which spatially coincide with compact H$\alpha$ structures. 
To estimate the star formation rate from our data, the H$\alpha$ map has been calibrated into physical units by measuring the intensity within circular profiles extracted from the FPI map and the calibrated H$\alpha$ continuum-subtracted direct image. The total H$\alpha$ flux within the ring of Hoag's Object is $(2.1\pm0.2)\times 10^{-14}$ erg s$^{-1}$ cm$^{-2}$ which corresponds
to a total luminosity of $(7.7\pm0.8) \times 10^{40}$ erg s$^{-1}$. 

Unfortunately, we cannot fit the spectrum of the continuum light of the ring with more complicated stellar population models due to its low quality.
A detailed imaging study of individual knots is also problematic due to the significant difference in spatial resolution between the HST and H$\alpha$ images. 
Alternatively, the photometry of the ring can be used to recover its stellar properties by modeling the continuum light and the blue star clusters as two distinct stellar populations.
For simplicity, we consider a mixed stellar population formed in two instantaneous bursts and use the stellar population synthesis models of Bruzual \& Charlot (2003) with a Salpeter (1955) initial mass function.
To get a better handle on the stellar populations we make use of the Sloan Digital Sky Survey (SDSS) $ugriz$ and Galaxy Evolution Explorer (GALEX) near-UV (NUV) photometry. The colours of the ring, measured after subtracting the underlying light of the core, are $u-g=1.15\pm0.12$, $g-r=0.43\pm0.05$, $r-i=0.05$, $i-z=0.78\pm0.23$ and $\mbox{NUV}-g=2.26\pm0.12$. Note that the errors reflect differences in photometric measurements in two SDSS imaging runs.
In addition, the number of ionizing Lyman-continuum photons, which is given by the models, is estimated from the H$\alpha$ flux under some simple assumptions (see Finkelman et al.\ 2010 for more details). 
The data are best-fit by $\sim$1 Gyr old and $\lesssim$10 Myr stellar populations with 2.5 and 0.2 $\mbox{Z}_\odot$, respectively. The mass fraction of young-to-old populations is about 10\%. Although the errors on the results are not large, we caution that five free parameters might not be sufficiently constrained by our limited data. The obtained values are therefore not reliable and should provide only a consistency check of conclusions based on our spectroscopic measurements.

\section{The structure of Hoag's Object}
\label{S:structure}
\subsection{The central core}
The presence of a blue detached ring, viewed face-on, around a red central core gives Hoag's Object its unique appearance. 
The core itself seems, however, to be a normal spheroid in all its photometric and kinematic properties (Schweizer et al.\ 1987). 
Whether the core is truly a bulge or an elliptical galaxy, a question raised by Schweizer et al.\ (1987), is probably irrelevant in light of our current understanding of galaxy formation processes.   
Classical bulges and ellipticals happen to have common general properties, including star formation history and chemical enrichment, 
which implies that they are in fact similar objects (Renzini 1999). 
In this view, ellipticals form as bulges in the early universe, but fail to grow or maintain a prominent disc for some reason. 
A different kind of bulge may form through the slow rearrangement of disc material and therefore have structure and kinematics resembling that of a disc (Fisher \& Drory 2008). Such dynamically-cold bulges are referred to as `pseudobulges' to distinguish them from the dynamically hot classical bulges (Kormendy \& Kennicutt 2004).

All the photometric evidence point to the core of Hoag's Object being a classical bulge.
It clearly shows a smooth and regular $r^{1/4}$ light profile with no features or inner structure, which is typical of an elliptical galaxy. 
Furthermore, the $B$-band central surface brightness and absolute blue magnitude of the core characterize it as a 'true elliptical' (see fig.\ 1 of Kormendy et al.\ 2009). The red colour gradient towards the centre of the core is also common of intermediate-luminosity ellipticals (Faber al.\ 1997).

The analysis of the HST images shows that the PA smoothly decreases from $\sim$75$^\circ$ at the centre to a value of $\sim$45$^\circ$ toward the edge of the core. 
Furthermore, the kinematic axis is only aligned with the photometric major axis toward the edge of the core.
Such an isophotal twisting is common in ellipticals where the light profile deviates from the $r^{1/4}$-law and is generally attributed to the misalignment of stellar orbits inside the bulge (Fasano \& Bonoli 1989). In this view, the variation in the PA and ellipticity along the $r^{1/4}$ core of Hoag's Object probably arise from its intrinsic triaxial  structure.

Do the kinematic properties of the core also appear to be characteristic of a classical bulge?
To obtain the central velocity dispersion we used the deepest spectral measurements (along PA=45$^\circ$, which also has the best seeing) and averaged the data inside $r<r_e=2.8$ arcsec. The calculated effective velocity dispersion of $\sigma=151\pm5$ km s$^{-1}$ is in agreement with the Schweizer et al.\ (1987) value of $\sigma=154\pm6$ km s$^{-1}$. The maximum observed rotation velocity along the major axis is $V=70 \pm 8$ km s$^{-1}$ at $r=4$ arcsec, which should be taken as a lower limit since the galaxy is seen in projection.
To further investigate the observed kinematics of the core we fitted the MPFS data adopting the `tilted-ring' method (Begeman 1989),
briefly described below and also in Moiseev et al. (2004). 
At first, we fixed the position of the kinematic centre to be the symmetry centre of the velocity field and found that it coincides within 1 pixel with the centre of the continuum image. We then fixed the rotation centre at the centre of the continuum image. We fitted the observational points with a simple circular rotation model and derived the systemic velocity $V_{sys} = 12767 \pm 3$ km s$^{-1}$, the kinematic position angle PA=$41 \pm 5^\circ$  and the equatorial plane inclination to the line-of-sight $i = 19 \pm 5 ^\circ$. We then divided the velocity fields into concentric elliptical annulus of one arcsec width with fixed values of $V_{sys}$, PA and $i$, measured the radial velocity as a function of azimuthal angle in the plane of the galaxy, and determined the mean rotation velocity $V_{rot}$ as a function of galactic radius.
As demonstrated in Fig.\ \ref{fig_rc}, the model rotation velocity is rising up to $180$ km s$^{-1}$ at the distance $r=6$ arcsec ($\simeq$2$r_e$) which is the limit of our kinematic measurements for the core.

The derived morphological and kinematical properties slightly deviate from the fundamental plane for ellipticals in log($r_e$), $\left\langle \mu_e \right\rangle $ and log$\sigma$ but are within the 3-$\sigma$ scatter (see Dressler et al.\ 1987; Faber et al.\ 1989; Reda et al.\ 2005).
We also placed the core on the anisotropy diagram, which relates the ratio of the ordered and random motion ($V/\sigma$) in a galaxy to its observed ellipticity. The $V/\sigma$ value measured for the core is $\sim$0.46, and when applying the inclination correction it increases to $\sim$1.2. 
Given that the ellipticity is less than 0.03 at $r_e$ indicates that the core is an extremely fast rotator (see fig.\ 9 in Cappellari et al.\ 2007). 

\subsection{The ring}
H$\alpha$ emission is detected along the entire ring, at least to the limit of the luminous stellar component, as shown by Fig.\ \ref{kinematic_figures}.
Adopting the titled-ring approximation we fitted the observational data with a simple circular rotation model and derived the systemic velocity $V_{sys}=12761\pm4$ km s$^{-1}$, the kinematical position angle PA$=43 \pm 3 ^\circ$ and the equatorial plane inclination to the line-of-sight $i=18 \pm 4 ^\circ$.
The fit was limited for regions inside $18<r<28$ arcsec since at smaller radial distances the velocity field is slightly perturbed (see below).
As shown on the left panel of Fig.\ \ref{fig_models} this simplified 'flat disc' model fits well the ionized gas velocity field with relatively minor velocity residuals.

\begin{figure}
\includegraphics[width=0.5\textwidth]{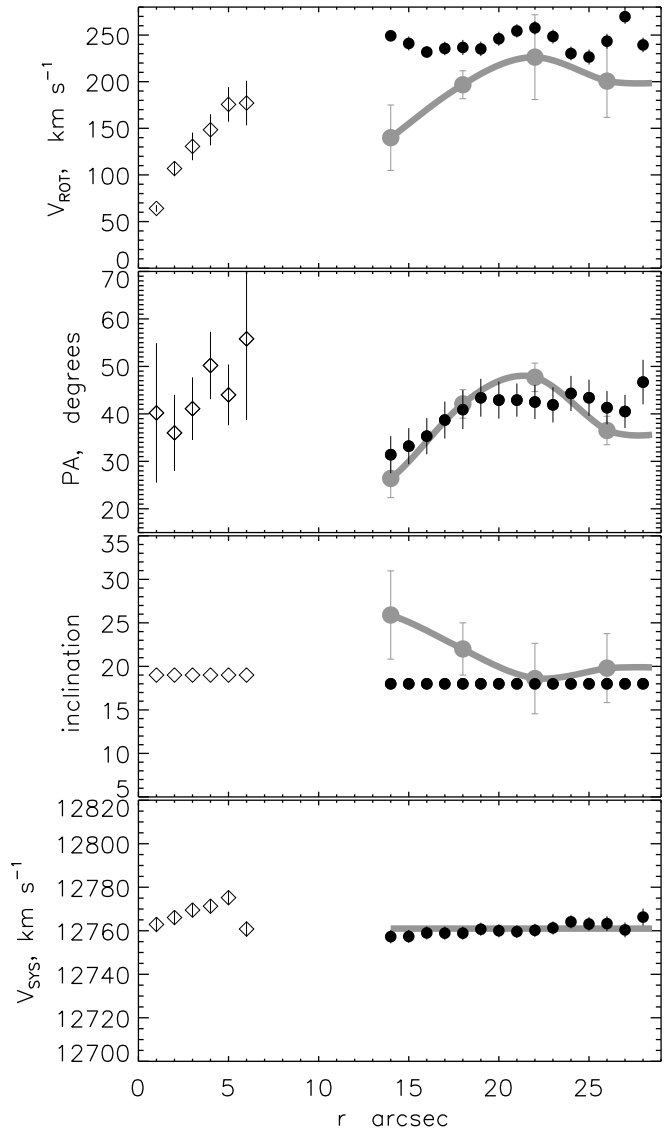}
\caption{Radial variations of kinematic parameters characterizing the velocity field of the ionized gas. From top to bottom: rotation velocity, position angle, inclination and systemic velocity. The filled circles and open diamonds represent the ``tilted-ring'' method results for the ionized gas (FPI data) and stars (MPFS data), respectively. The thick gray line shows the warp model fitted parameters.}
\label{fig_rc}
\end{figure}

A second model, labeled `tilted-rings' on Fig.\ \ref{fig_models}, used three free parameters for fitting the observed velocities in each elliptical annual: PA, $V_{rot}$ and $V_{sys}$. Since this model is indeterminate to individual measurements of the inclination of narrow, nearly face-on rings, the value of $i$ was kept fixed.
Fig.\ \ref{fig_rc} shows the radial variations of the model parameters whereas the model velocity field together with the residuals map are displayed on the lower panel of Fig.\ \ref{fig_models}. 
The ionized gas rotation curve is nearly flat with $V_{max}=260$ km s$^{-1}$ and the systemic velocity agrees with our fit to the stellar core. The observed PA twist at $r=14-18$ arcsec might be related to non-circular motions in the plane of the ring or with circular rotation in inclined orbits. To further investigate the nature of these orbits we fitted the velocity field with a `warp disc' model similar to that described by Moiseev (2008). Instead of using analytic functions, as in Moiseev (2008), the PA, $i$ and $V_{rot}$ radial variations were determined with four fixed radial nodes and were interpolated to all radii using spline functions. The resulting two-dimensional model was then smoothed to the spatial resolution of FPI data and fitted to the observed velocity field by a $\chi^2$-minimization procedure. This approach fundamentally differs from the tilted-rings model since the model is first constructed for the entire velocity field and only then is fitted to the observations. 
The thick gray line in Fig.\ \ref{fig_rc} shows the radial variations of the best-fit warp disc parameters, whereas the model ionized gas velocity field is presented in the right panel of Fig.\ \ref{fig_models}. 
The residual velocities for the three models are typically less than $20$ km s$^{-1}$ with the warp disc model best describing the observed velocities at  $r<18$ arcsec. However, as we argue below, the resulting parameters of the warp disc appear problematic and unreliable.

If Hoag's Object has a large HI disc, it follows that we can trace only a small part of it where ongoing star formation produces significant H$\alpha$ emission. 
We therefore must extrapolate the observed radial variation of the PA to the centre. 
Our extrapolation of the inward gas rotation curve significantly disagrees with any reliable mass distribution along the central core or with the stellar rotation derived from the MPFS data.  
The inner orbits in a disc are expected to be more stable than the outer orbits due to the increasing surface density and shorter relaxation times closer the galaxy centre. Although inner strong warps and polar discs are known to exist, these are typically associated with a compact or circumnuclear structure rather than with a large-scale disc (e.g., Bettoni, Fasano \& Galletta 1990, Corsini et al.\ 2003, Sil'chenko \& Afanasiev 2004). Large-scale polar structures in early-type galaxies can form around a triaxial or tumbling gravitational potential, but tend to show visible signs of inner star formation or dust lanes crossing the stellar core (see Sparke 1996; Sparke et al.\ 2009; J\'{o}zsa et al.\ 2009).
Moreover, warped discs are often inclined to the main plane of the stellar core where the gas lies outside the galaxy plane due to a specific orientation of the angular momentum of accreting matter (e.g., van Driel et al.\ 1995; Moiseev 2008). This is in contradiction with the kinematical alignment between the stellar core of Hoag's Object and its ionized gas component at distances $r>18$ arcsec.

We therefore dismiss the warped disc hypothesis and suggest that non-circular motions provide a more realistic explanation of the ionized gas dynamics. 
Such motions may be related with radial motions caused by a non-axisymmetric potential, or more likely with streaming motions along a spiral pattern, as observed in Hoag's Object optical images.
We conclude also that the spiral pattern is unlikely an helical structure which only by chance coincides with the core plane and is viewed nearly face-on.
First, as demonstrated in Fig.\ \ref{fig_models}, even the warped disc model cannot reproduce all the features of the velocity field, implying that at least part of the motion is non-circular.
Second, galaxies with helical structures always present extended  high- or low-contrast features (shells and envelopes, see Makarov, Reshetnikov \& Yakovleva 1989; Sparke et al.\ 2008; Sparke et al.\ 2009) whereas we could not detect any such features outside the ring of Hoag's Object down to very low surface brightnesses (see Fig.\ \ref{fig:deep_BTA}). 
In addition, the light distribution along the ring of Hoag's Object varies only little relative to the brightness fluctuations observed along helical structures in other galaxies.

\begin{figure*}
 \includegraphics[width=\textwidth]{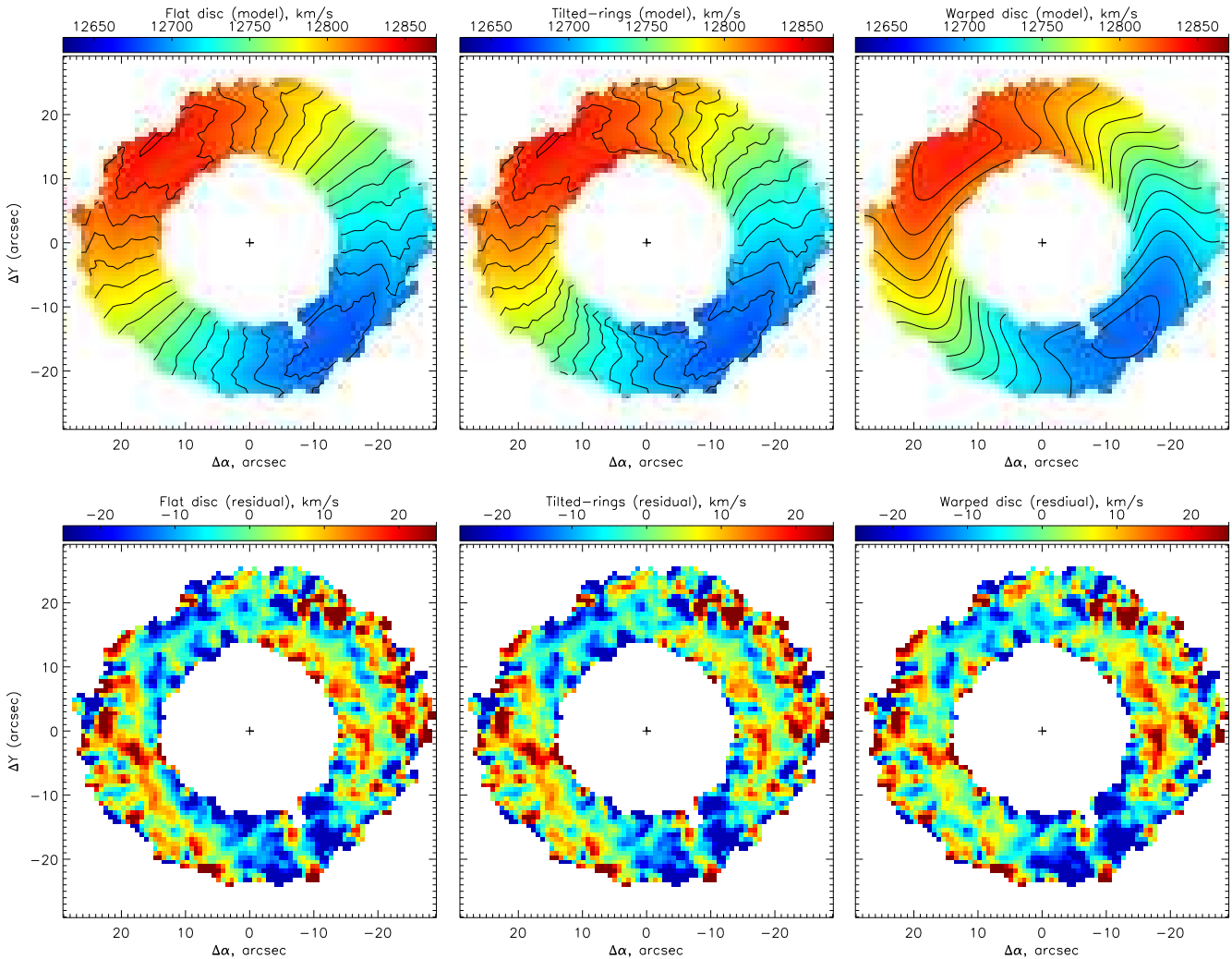}
\caption{Modeling of the ionized gas velocity field. The upper panels represent three different models while the lower panels represent the models corresponding residual maps, constructed by subtracting the models from the observed data.}
\label{fig_models}
\end{figure*}

\section{The formation of Hoag's Object}
\label{S:hypothesis}
The detailed photometry and kinematic study of Hoag's Object provides reliable clues concerning the nature of this peculiar galaxy.
We use the new data to test old formation hypotheses of Hoag's Object and provide an alternative suggestion.  

\subsection{Review of the old hypotheses}
\subsubsection{Bar instability}
After establishing the extragalactic nature of Hoag's Object, several scenarios were put forward in an attempt to account for the structure of the galaxy.
An explanation by Brosch (1985) suggested that the ring could be related with the `resonance hypothesis', i.e., formed around the core due to a dynamical Lindblad resonance in response to internal non-axisymmetric perturbations. 
In this view, the ring was formed in a similar manner as rings in barred spiral galaxies by slow internal evolution and without requiring galaxy interactions (see also Wakamatsu 1990); a bar transfers angular momentum to the outer disc, thereby driving gas outwards from the centre where it collects into an outer ring near resonance (Kormendy \& Kennicutt 2004). 
This process is expected to instantaneously trigger star formation throughout the ring, indicating that the observed colours of the ring are of a stellar population few Gyr old. 
However, his failure to detect a bar led Brosch to conclude that Hoag's Object is truly a S0 galaxy where the disc suffered a strong bar instability, leaving an extremely faint bar. 

New evidence about the nature of Hoag's Object emerged with the detection by Schweizer et al.\ (1987) of an HI component significantly more massive than the upper limit determined by Brosch (1985). The two-horned HI profile was interpreted as typical of discs with a flat or slightly rising rotation curve rather than with a thick shell. 
The presence of such a massive HI component is expected to stabilize a disc against bar instability (Sellwood 1981). This was given by Schweizer et al.\ (1987) as a strong argument against the Brosch (1985) hypothesis.

The photometric and kinematic properties of the core, together with their failure to detect a bar or any evidence for a discy component, led Schweizer et al.\ (1987) to conclude that Hoag's Object is a `normal spheroid for its luminosity'. 
The new photometric and kinematic data presented here leave no doubt that the core of Hoag's Object exhibits no disc-like fine structure in its central regions. Furthermore, the $\gtrsim$10 Gyr old stellar population of the core argues against gas inflow and a late star formation episode, which are expected during the evolution of a bar. Thus, the core can be safely characterized as a true elliptical surrounded by a ring (see also Sandage \& Bedke 1994; Finkelman \& Brosch 2011).

\subsubsection{Major accretion event}
Schweizer et al.\ (1987) discussed also the hypothesis that Hoag's Object resulted from a direct collision or close encounter of two nearby galaxies.
The authors found that the core and the ring have similar recession velocities and concluded that they must form a single object. They therefore ruled out an older hypothesis that Hoag's Object is a classical collisional ring seen with its intruding companion superimposed (Hoag 1950; O'Connell, Scargle \& Sargent 1974). However, as pointed out by Appleton \& Struck-Marcell (1996), the peculiar configuration of the core and the ring can be explained if Hoag's Object is viewed at an oblique angle to our line-of-sight and exactly as the intruding companion reaches its furthest distance from the target, where it is at rest relative to the ring.
This possible combination of rare circumstances can be dismissed since our data strongly indicate that the system has nearly reached equilibrium. Since such a relaxed structure is also seen in the Hoag-type galaxy UGC 4599, we conclude that the collisional ring scenario is not responsible for forming genuine Hoag-type galaxies (Finkelman \& Brosch 2011).

Schweizer et al.\ (1987) put forward a new hypothesis that Hoag's Object formed in a `major accretion event 2-3 Gyr ago' by mass transfer or merger, in a mechanism similar to that forming polar ring galaxies. 
The authors emphasized that they could not detect any tidal tail, ripple signatures or merging signature which are expected if an interaction-driven event took place recently. They also mentioned that the bright knots in the ring are aligned in a narrow ring which is off-centre with respect to the main ring, implying that the ring has not yet settled since the event took place. 

Do our observational data agree with the Schweizer et al.\ (1987) scenario?
Closely examining the detached ring structure constrains the age of the presumed accretion event.
The stellar population of the ring was derived in Section \ref{S:pop} by fitting the observational data with SSP models. 
The spectral stellar population analysis and colours of the ring indicate the presence of a $\lesssim$2 Gyr old stellar population.
There are also clear signatures that star formation is currently taking place in the ring.

The major accretion event can also be tested by looking for signatures of  extended fine structure or of morphological distortion and asymmetries.
The deep BTA optical image reveals no faint tidal tails nor low surface brightness features, supporting the idea that any presumed event should have occurred at least 2-3 Gyr ago (e.g., Hibbard \& Mihos 1995; Duc et al.\ 2011). For the gas to settle in ordered orbits, as demonstrated by the two-horned HI profile and by the ionized gas velocity map, several orbits are necessary. The orbital period at the radius of the ring is $\sim$0.4 Gyr which implies that the gas required at least $\sim$1 Gyr to settle in circular ordered orbits, depending on the true size of the HI disc.
To confirm that the atomic gas is in a regularly rotating HI disc or ring, and to detect whether it connects with unsettled HI structures, such as tails and clouds, requires high-resolution 21-cm mapping.

These indications seem to put strong limits on the age of the ring of Hoag's Object, and are supposedly consistent with the Schweizer et al.\ (1987) hypothesis.
However, several other considerations might prove difficult to explain with a major accretion scenario.
The total blue light coming from the ring is about $3.3\times10^9 \, \mbox{L}_\odot$. A $\sim$1-3 Gyr old stellar population with $\mbox{Z}=0.4 \, \mbox{Z}_\odot$ (see Fig.\ \ref{fig:age}) has a predicted mass-to-light ratio of about 1-3 (Bruzual \& Charlot 2003) which indicates that the total mass of stars in the ring is larger than 3$\times 10^9 \, \mbox{M}_\odot$. 
Examining the archived GALEX NUV image of Hoag's Object, and following the star formation rate calculation by Salim et al.\ (2007), we estimate the star formation rate within the entire ring to be $\sim$0.7$\, \mbox{M}_\odot$ yr$^{-1}$. 
Adopting the Kennicutt (1998) relation, we estimate the star formation rate from the H$\alpha$ luminosity to be $0.7\pm0.1 \, \mbox{M}_\odot$ yr$^{-1}$, in agreement with the UV estimate. Such a continuous star formation rate would have formed the ring within more than 4 Gyr. 

Adding the total mass of stars in the ring to the observed HI mass implies that the initial HI mass in the disc was well over $10^{10} \, \mbox{M}_\odot$. Such an amount of HI gas is more than twice that of the Milky Way and is at least one order of magnitude larger than the typical HI content of dwarf irregular galaxies (Matthews, Gallagher \& Littleton 1993; Matthews \& Gallagher 1996). 
This implies that the gas could not have been accreted from a single event by a passing gas-rich dwarf companion (see also Galletta, Sage \& Sparke 1997; Sparke \& Cox 2000, Sancisi et al.\ 2008). The accretion of a large number of small neighbours also seems unlikely, considering the low-density environment of Hoag's Object and that different companions would interact at random angles to Hoag's Object. 
However, note that our environmental study is based only on objects with measured redshift included in NED. Previous studies of gas-rich early-type galaxies have demonstrated that nearby companions with detectable atomic gas are often missed by optical catalogues or show no sign of an optical counterpart (e.g., Serra et al.\ 2006; Struve et al.\ 2010). In some cases, radio observations reveal also signs of interactions with neighboring galaxies, such as HI bridges or tails (e.g., Finkelman \& Brosch 2011). 
Exploring the HI environment at high sensitivity is therefore essential to complement our optical study.

An equal-mass merger of two massive galaxies could account for the high HI content and reproduce the observed stellar distribution characteristic of an elliptical galaxy (Mulchaey \& Zabludoff 1999; Reda et al.\ 2004, 2007; Bournaud, Jog \& Combes 2007). 
In this scenario, at least a fraction of the initial gas is expected to lose angular momentum
and fall toward the centre of the remnant, subsequently triggering star formation and probably nuclear activity.
Some of the gas in the two merging galaxies can retain most of its angular momentum and spread at large radii, and at a later stage be re-accreted to settle in an extended, massive disc around the merger remnant (Serra et al.\ 2008 and references therein). 
A merger-induced central starburst would have used a large fraction of the cold gas of the core, which can explain the low amounts of atomic and molecular gas reported in the literature. 
A robust analysis of the SDSS spectra of Hoag's Object using the {\small VESPA} code (VErsatile SPectral Analysis, Tojeiro et al.\ 2007) hints at such a late event in the evolution of the core.
{\small VESPA} provides detailed star formation histories for the SDSS DR5 galaxies obtained by fitting all of the available absorption features, as well as the shape of the continuum, with synthetic models. The {\small VESPA} analysis of the core shows that it formed more than 9 Gyr ago, but contains also a small fraction (about 3\% in mass) of $\sim0.5$ Gyr stars. 
However, we emphasize that our deep spectroscopic measurements fit very well a single stellar population of $\gtrsim$10 Gyr (see Fig.\ \ref{f:spectra}). 

The imprint of a late starburst might be difficult to detect by optical means.
Therefore, more conclusive constraints on the galaxy merging history should be found by fitting a detailed UV-to-IR spectral energy distribution of the galaxy with composite stellar population models. 
For instance, 1-Gyr-old instantaneous starburst might leave an observational signature in the mid-IR (Vega et al.\ 2010).
Recently formed massive stars could be traced by their NUV light. The NUV-optical colour would become bluer even if the core of Hoag's Object has suffered a small amount of recent residual star formation. Using SDSS and GALEX archival data we measure $\mbox{NUV}-r \approx 5.1$ which puts Hoag's Object below the red sequence limit but does not necessarily require a recent star formation episode (Hern\'{a}ndez \& Bruzual 2009). 

Finally, we turn to test the major accretion hypothesis from a kinematical point of view.
The relaxed kinematics of the core indicate also that it could not have formed too recent. 
The specific kinematical properties of the merger remnant depend strongly on the details of the merging
process and specifically on the relative angular momenta of the two merging galaxies (Toomre 1977; Mihos \& Hernquist 1996, Barnes 2002; Bournaud, Jog \& Combes 2005; Di Matteo et al. 2007). 
Numerical simulations show that fast rotators can maintain the angular momentum content from the progenitor disc galaxies, depending on their mass ratio.
However, fast rotating E/S0 galaxies appear to resemble the progenitor galaxies more than the simulated merger remnants, which raises doubts about the importance of mergers in forming this type of galaxies (Jesseit et al.\ 2009).
In particular, the formation of a round, extremely fast rotating galaxy such as Hoag's Object appears to pose a serious challenge to the standard view of galaxy mergers. 
State-of-the-art high-resolution numerical simulations of binary mergers of disc and early-type galaxies found that most of the fast rotators produced in major mergers have intermediate flattening, with ellipticities between 0.4 and 0.6 (Bois et al.\ 2011). 
The majority of these simulated galaxies also have a bar. 

To conclude, the physical properties of the ring are consistent with (although not supportive of) the Schweizer et al.\ (1987) hypothesis. However, the major accretion event is, to say the least, inconsistent with our kinematical observations and stellar population modeling of the core. We therefore deem this scenario unlikely, although further observations are required to disprove it.
\subsection{A new hypothesis}
\subsubsection{Formation of the core}
Hoag's Object is located in a very low-density environment with the closest companion with known redshift $\sim$3 Mpc away.
Different possible formation scenarios have been proposed for isolated ellipticals by observational studies, but their nature remains debated and a detailed study of their properties is still required (Niemi et al.\ 2010).
If the core of Hoag's Object did not form in a major merger it could be the result of a monolithic gas collapse.
According to this model, stars form during a rapid dissipative collapse while the gas sinks to the centre of the forming galaxy and is chemically enriched by evolving stars (Larson 1974, 1975; Carlberg 1984; Arimoto \& Yoshii 1987).
Such a scenario naturally explains the smooth $r^{1/4}$ light distribution of the core.

The metallicity of the central component of Hoag's Object can provide an important clue to the formation of the core.
A major merger of two massive progenitors is expected to produce a more metal-rich
stellar population and shallower metallicity gradient than the predicted gradients of dissipative collapse models.
The gas around the core must have the same origin as the gas from which stars formed. Thus, a continuous star formation process in the ring would lead to a chemical enrichment of the young stars of the ring with respect to the stars in the core. 

To state definite conclusions about the metallicity and age gradients, and break the age-metallicity degeneracy, we determined them
simultaneously in our model fit. 
We estimate the metallicity gradient along the core of Hoag's Object to be $\mbox{d([Z/H])/dlog}r=-0.4$ dex per dex. This value does not allow distinguishing with confidence between the monolithic dissipative collapse picture and a major merger (see Kobayashi 2004; S\'{a}nchez-Bl\'{a}zquez et al.\ 2006; Baes et al.\ 2007; Spolaor et al.\ 2009 and references therein). However, we note that the metallicity profile of the core is probably smeared by seeing effects, thus the true metallicity gradient is probably steeper, which would be more consistent with the monolithic collapse hypothesis.

Explaining the fast rotation of the core by the typical anisotropy of an oblate (or mildly triaxial) spheroid would require it to be extremely flattend with intrinsic ellipticity close to 1, which seems very unlikely. This implies that, since its formation, Hoag's Object gained additional angular momentum from an external source. Accretion of gas from outside could have increased $V/\sigma$ significantly assuming that the additional gas forms stars (Emsellem et al.\ 2007). Such a process should preferentially occur in gas-rich environments at high-redshifts or in low-density regions (Emsellem et al.\ 2011). 

\subsubsection{Formation of the ring}
Although an extended HI structure could have formed around the central core in a gas-rich merging system (Barnes 2002), we showed that the large gas content was more likely accreted early in the galaxy evolution. 
Accretion of gas from the intergalactic medium (IGM) during a prolonged, dynamically-quiet
phase can explain both the flow of substantial amounts of gas onto a pre-existing early-type galaxy (Kere\v{s} et al.\ 2005; Serra et al.\ 2008; Dekel, Sari \& Ceverino 2009; Bournaud \& Elmegreen 2009) and the extremely high specific angular momentum of the core.
If part of the gas is not shock heated to the virial temperature of the halo and reaches a maximum temperature below $10^5$ K, then it can cool to the neutral atomic state on a timescale of several Gyr while forming a dilute and extended atomic gas structures. The amount of gas accumulated by this 'cold' accretion process depends on the halo mass and on environmental properties.

Observational evidence for the theoretical picture of IGM accretion are still poor (e.g., Serra et al.\ 2008; Steidel et al.\ 2010). Deep HI observations of nearby spiral galaxies revealed large quantities of extraplanar cold gas at large vertical distances which may provide important clues to the roles of star formation and accretion (Heald et al.\ 2010 and references therein). While most of the gas in these halos is probably related to star formation-driven disc-halo flows (e.g., Bregman 1980), a small fraction might be accreted from the IGM. This is hinted by the  presence of large HI streams and small, counter-rotating HI clouds found in these gas halos (Morganti et al.\ 2006; Oosterloo, Fraternali \& Sancisi 2007). 
Accretion from the cosmic filament web could also account for the formation of polar structures around early-type galaxies (Macci\'{o} et al.\ 2006; Brook et al.\ 2008). In fact, the formation of a central spheroid surrounded by a massive, low-luminosity disc of high HI content seems to be more consistent with cold accretion processes rather than with galaxy interactions (Spavone et al.\ 2010). Furthermore, a growing line of evidence for cold gas accretion in isolated galaxies, including such with polar components, suggests that this is a realistic way of building up galaxies in underdensed environments (Stanonik et al.\ 2009; van de Weygaert et al.\ 2011).

The two-horned shape of the 21-cm line profile of Hoag's Object implies that it lacks a prominent gas halo. This is not surprising since the core of Hoag's Object is an isolated elliptical galaxy with no active star formation. 
The ionized gas velocity curve of the ring implies that Hoag's Object has a halo mass of $M_{\mbox{halo}}=Rv^2/G\approx 3\times10^{11} \, \mbox{M}_\odot$ within a radius of 20 kpc. 
While such a halo mass is probably not large enough to accrete in the early Universe most of the $10^{10} \, \mbox{M}_\odot$ HI content (Kere\^{s} et al.\ 2005; Serra et al.\ 2006; Finkelman \&  Brosch 2011) we note that our calculation sets only a lower limit on the mass, which would be larger by an order of magnitude if Hoag's Object is indeed surrounded by an extended HI disc.

The cold gas accretion mechanism for disc formation predicts rather low metallicities of $\mbox{Z}=0.1 \, \mbox{Z}_\odot$ (Dekel \& Birnboim 2006; Ocvirk, Pichon \& Teyssier 2008; Agertz, Teyssier \& Moore 2009). Subsequent enrichment due to ongoing star formation activity during the last few Gyr could account for metallicities as high as those of outer discs of bright spiral galaxies (Bresolin et al.\ 2009; Spavone et al.\ 2010). 
Metal enrichment influenced by the stellar evolution of the old core itself is expected to produce a metallicity gradient along the ring. Unfortunately, all of these cannot be tested with our spectroscopic data due to the large uncertainty in the metallicity measurements.

\subsubsection{Evolutionary path}
The significant HI gas reservoir of Hoag's Object is not unusual for early-type galaxies (Morganti et al.\ 2006).
In particular, recent studies revealed extended, low-level star formation activity in the form of outer rings and spiral arms around a surprisingly high fraction of these galaxies (Thilker et al.\ 2007; Donovan et al.\ 2009; Salim \& Rich 2010; Thilker et al.\ 2010; Finkelman \& Brosch 2011). In most cases the star formation is probably fueled by newly acquired gas from minor gas-rich mergers or from the IGM. Such late, large-scale star formation could potentially transform the galaxy while building its stellar disc. 

Hoag's Object may represent an extreme form of this class of galaxies as an evolved stage of an elliptical galaxy formed early in cosmic time (Marcum, Aars \& Fanelli 2004). The infalling primordial or intergalactic gas probably evolved only very slowly for several Gyr until eventually settling into a stable orbit around the forming or pre-existing central bulge (Schwarz 1981; Pearce \& Thomas 1991; Finkelman \& Brosch 2011).
Hoag's Object would be a particularly promising case to investigate if gas accretion plays a more significant role in low-density environments (Oosterloo et al.\ 2010). 

Deep HI imaging is necessary to map the extended halo of cold gas clouds around Hoag's Object, or to detect tidally-disrupted dwarf companions.
The exact details of gas accretion depend on the presence of X-ray gas in the galactic halo, which could ionize at least part of the cold gas before it reaches the disk (e.g.\ Nipoti \& Binney 2007).
The detection of AGNs in low star formation environments, such as elliptical galaxies, might account also for a delayed, extended star-forming phase.
While no clear sign of recent nuclear activity is detected in optical spectra of the nucleus of Hoag's Object, AGNs are commonly associated with star formation quenching during their initial growth phase (e.g., Croton et al.\ 2006; Salim \& Rich 2010).
A detailed mapping of Hoag's Object in the X-ray and the far-UV wavelengths could therefore contribute to our understanding of the interplay between the various gas phases and of the role of AGN feedback processes in this galaxy.

If the gas is distributed over an extended disc around Hoag's Object it would be relatively dilute.
Detailed VLA observations of the Hoag-type galaxy UGC 4599 reveal an extended $\sim$100 kpc HI disc where the dilute gas is denser along spiral arms and a ring which coincide with the optical ring (Dowell 2010). Without detailed HI maps of Hoag's Object we cannot state whether or not the HI distribution exceeds the apparent edge of the ring at about 25 kpc and shows similar features.
Given the measured ionized gas velocity and a typical velocity dispersion of about 7 km s$^{-1}$, and following Serra et al.\ (2006), we estimate the critical surface density for star formation to be about $5.6\, \mbox{M}_\odot$ pc$^{-2}$. However, assuming the HI distribution has a central hole and is associated only with the ionized gas ring, the average surface density of the neutral gas cannot be larger than $\sim$4.8$\, \mbox{M}_\odot$ pc$^{-2}$, i.e., below the star formation threshold.
This can be explained if the canonical star formation law we adopted cannot be extrapolated to the outer parts of (disc) galaxies.

The presence of spiral features (and warps) in the outer regions of extended HI discs around galaxies is generally related to bars or oval discs (Sancisi et al.\ 2008), which are not present in genuine Hoag-type objects. 
Schweizer et al.\ (1987) mentioned that an elongated triaxial core could, in principle, successfully reproduce the ring of Hoag's Object without requiring a bar (see also Wakamatsu 1990). 
Although this idea was dismissed due to what appeared to the authors as a rotationally symmetrical core, we interpret the variation in PA and ellipticity as evidence for triaxiality. 
An oval core structure, or other weakly non-spherical potentials, such as a passing companion, could continuously generate spiral waves in the diffuse HI disc (see Buta 1986, Buta \& Combes 1996; Rautiainen \& Salo 2000; Bekki \& Freeman 2002; Tutukov \& Fedorova 2006). 
These, in turn, could condense the gas and trigger star formation (Sancisi et al.\ 2008; Bush et al.\ 2010; Khoperskov et al.\ 2010).
The twisting of the spiral structure by the differential rotation could lead to the formation of a long-lived ring, provided that the star formation is not too intense. Such an event would account for the gap in the light distribution between the core and the ring (Schwarz 1981; Pearce \& Thomas 1991) as well as for the moderate metallicity and the old age of the core.

\section{Conclusions}
\label{S:conclusions}
We analyzed here new observations of Hoag's Object, consisting of multiband photometry with the HST/WFPC2 and long-slit and 3D spectroscopic data observations with the Russian 6-m telescope. 
We showed that Hoag's Object has a slightly elongated bulge which exhibits the typical morphology of an elliptical galaxy and the dynamics of an extremely fast rotator.
A variation of the PA and ellipticity implies that the isophotes are intrinsically twisted, i.e., the core is not axisymmetric.
The PA gradually decreases from the centre outwards to the true kinematical PA, which provides a strong evidence that the system had settled to its current configuration long time ago and is now in equilibrium.  
Considering the age of the host galaxy and the massive, extended HI content we concluded that the core formed more than $\sim$10 Gyr ago and that the HI disc probably formed a short time after. 

The ring itself shows extended H$\alpha$ emission, including also several clumps and knots where the emission is more intense.
The rotation of the ionized gas ring seems to follow the circular motion of the cold atomic hydrogen component. Non-circular velocity residuals of $\lesssim$20 km s$^{-1}$ are probably related with the optical spiral pattern across the ring. 
Having failed to detect a bar or a central discy component, we suggest that a nonaxisymmetric potential, presumably the triaxial core, 
induces a spiral motion along the HI disc outside the core, which leads, in turn, to the continuous formation of stars. This could also explain the prominent gap between the core and the ring.  

Considering all pieces of observational evidence we conclude that accretion of gas from the IGM onto a pre-existing elliptical galaxy best explains both the peculiar structure and kinematics of Hoag's Object.
However, a better understanding of this process is needed to put firm constraints on the expected properties of post-accretion systems.
A detailed HI study is required to investigate the surroundings of Hoag's Object and its cold gas disc and help to further understand the nature of this galaxy.

\subsection*{Acknowledgments}
We are grateful for the generous allocation
of observing time to this project by the Time Allocation
Committee of the Special Astrophysical Observatory.
The 6-m SAO RAS telescope is operated under the financial
support of the Ministry of Science and Education (registration
no.\ 01-43). We thank Dr Alexander Burenkov for assisting in the observations with the MPFS.
We also thank the Hubble Heritage team at the Space Telescope Science Institute for the images of Hoag's Object.

AM and IK thank the Russian Foundation for Basic Research (project no.\ 09-02-00870, 07-02-00005 and 08-02-00627). 
AM is also grateful to the 'Dynasty' Fund and the Russian Federal Program 'Kadry' (contract no.\ 14.740.11.0800).

This research has made use of the NASA/IPAC Extragalactic Database
(NED) which is operated by the Jet Propulsion Laboratory,
California Institute of Technology, under contract with the
National Aeronautics and Space Administration. 

Based on observations made with the NASA/ESA HST, obtained from the data archive at the STScI. 
STScI is operated by the association of Universities for Research in Astronomy, Inc.\ under the NASA contract NAS 5-26555.

Based on observations made with the NASA GALEX.
GALEX is operated for NASA by the California Institute of Technology under NASA contract NAS5-98034.

Funding for the SDSS and SDSS-II has been provided by the Alfred P.\ Sloan Foundation, the Participating Institutions, the National Science Foundation, the U.S.\ Department of Energy, the National Aeronautics and Space Administration, the Japanese Monbukagakusho, the Max Planck Society, and the Higher Education Funding Council for England. The SDSS Web Site is http://www.sdss.org/.

\end{document}